\newif\ifarxiv
\arxivtrue

\ifarxiv
\documentclass[a4paper, preprint]{elsarticle}
\else
\documentclass[a4paper, review]{elsarticle}
\fi

\usepackage[utf8]{inputenc}
\usepackage[T1]{fontenc} 
\usepackage{microtype} 
\usepackage{todonotes}
\usepackage{hyperref}
\usepackage{lmodern}

\usepackage{amssymb}

\usepackage{lineno}

\usepackage{cleveref}

\usepackage{fancyvrb}
\usepackage{color}
\input{listings/output/pygmentsDefs}
\expandafter\def\csname PY@tok@err\endcsname{} 

\usepackage{siunitx}
\usepackage{subcaption}

\newcommand{\walberla}{\textsc{waLBerla}}

\definecolor{color1}{rgb}{0.00392156862745098, 0.45098039215686275, 0.6980392156862745}
\definecolor{color2}{rgb}{0.8705882352941177, 0.5607843137254902, 0.0196078431372549}
\definecolor{color3}{rgb}{0.00784313725490196, 0.6196078431372549, 0.45098039215686275}
\definecolor{color4}{rgb}{0.8352941176470589, 0.3686274509803922, 0.0}
\definecolor{color5}{rgb}{0.8, 0.47058823529411764, 0.7372549019607844}
\definecolor{color6}{rgb}{0.792156862745098, 0.5686274509803921, 0.3803921568627451}
\definecolor{color7}{rgb}{0.984313725490196, 0.6862745098039216, 0.8941176470588236}
\definecolor{color8}{rgb}{0.5803921568627451, 0.5803921568627451, 0.5803921568627451}

\usetikzlibrary{fit}

\begin{document}

\begin{frontmatter}



\title{\walberla{}: A block-structured high-performance framework for multiphysics simulations}

\author[lss]{Martin~Bauer}
\author[lss]{Sebastian~Eibl}
\author[lss]{Christian~Godenschwager}
\author[lss]{Nils~Kohl}
\author[stuttgart]{Michael~Kuron}
\author[lss]{Christoph~Rettinger}
\author[lss]{Florian~Schornbaum}
\author[lss]{Christoph~Schwarzmeier}
\author[lss]{Dominik~Thönnes}
\author[lss]{Harald~K\"{o}stler}
\author[lss,cerfacs]{Ulrich~R\"{u}de}

\address[lss]{Chair for System Simulation, Friedrich--Alexander--Universität Erlangen--Nürnberg, Cauerstraße 11, 91058 Erlangen, Germany}
\address[cerfacs]{CERFACS, 42 Avenue Gaspard Coriolis, 31057 Toulouse Cedex 1, France}
\address[stuttgart]{Institute for Computational Physics, Universität Stuttgart, Allmandring 3, 70569 Stuttgart, Germany}

\begin{abstract}
Programming current supercomputers efficiently is a challenging task. 
Multiple levels of parallelism on the core, on the compute node, and between nodes need to be exploited to make full use of the system. Heterogeneous hardware architectures with accelerators further complicate the development process. 
\walberla{} addresses these challenges by providing the user with highly efficient building blocks for developing simulations on block-structured grids. 
The block-structured domain partitioning is flexible enough to handle complex geometries, while the structured grid within each block allows for highly efficient implementations of stencil-based algorithms. 
We present several example applications realized with \walberla{}, ranging from lattice Boltzmann methods to rigid particle simulations. 
Most importantly, these methods can be coupled together, enabling multiphysics simulations.
The framework uses meta-programming techniques to generate highly efficient code for CPUs and GPUs from a symbolic method formulation.
To ensure software quality and performance portability, a continuous integration toolchain automatically runs an extensive test suite encompassing multiple compilers, hardware architectures, and software configurations. 
\end{abstract}

\begin{keyword}
high-performance computing \sep multiphysics \sep lattice Boltzmann \sep rigid particle dynamics \sep adaptive mesh refinement \sep code generation
\end{keyword}

\end{frontmatter}

\ifarxiv
\else
\linenumbers
\fi


\section{Introduction}

Complex phenomena in the natural and engineering sciences are increasingly being studied with the help of simulations. 
A drastic increase in available computational power over the last few years now allows for large, highly resolved simulations.
Thus, Computational Sciences and Engineering (CSE) is emerging as a third fundamental pillar to complement theory and experiment \cite{keyes2013multiphysics,RuedeCSE2016}.
CSE as a new discipline aims at designing, analyzing, and implementing new efficient simulation methods on current high-performance computing (HPC) systems such that they can be applied to a wide variety of scientific and engineering problems in a robust, user-friendly, and reliable fashion. 
This requires software tailored to these particular needs.

\walberla{} is a modern open-source software framework that supports complex multiphysics simulations, and that is specifically designed to address the performance challenge in CSE: exploiting the full power of the largest supercomputers for a wide class of scientific research questions.

Here, performance and efficiency are crucial since simulation-based science often requires very fine spatial and temporal resolution to resolve all relevant physical effects. 
\walberla{} is carefully designed as a framework for massively parallel HPC systems, employing only fully distributed data structures \cite{godenschwager_framework_2013,feichtinger_2011}. 
Every process holds information only about local and adjacent data. 
Thus, the memory usage of a process does not depend on the total size of the simulation, making perfect scalability possible. 
Offering both pure Message Passing Interface (MPI)~\cite{MPI} and hybrid MPI/OpenMP parallelization, the underlying hardware can be used to its fullest potential.
Besides scalability, another major design goal of \walberla{} is a high node-level performance for all its compute kernels. 
A careful performance engineering process, together with automated performance tests, ensure sustainable performance across 
different compilers, operating systems, and hardware architectures.

HPC software usually has to be modified extensively in order to make full use of new hardware architectures.
This portability typically involves a lot of code duplication or at least similar code structure. 
To increase productivity by reducing duplication, \walberla{} employs code generation techniques to generate time-critical numerical kernels from a high-level, domain-specific formulation.

\walberla{}'s main focus are computational fluid dynamics simulations with the lattice Boltzmann method (LBM).
It therefore offers a wide range of state-of-the-art LBM models, together with a variety of utility and usability functionality.
In this regard, it is comparable to other LBM frameworks such as OpenLB~\cite{Heuveline2007,OpenLB}, Palabos~\cite{Lagrava2012,Palabos}, elbe~\cite{Mierke2018,elbe}, LB3D~\cite{Groen2011,schmieschek2017,LB3D}, HemeLB~\cite{Groen2013}, and Sailfish~\cite{Sailfish}.
Other LBM codes focus on HPC implementations of the method targeted at specific hardware architectures or a particular collision operator, like SunwayLB~\cite{Liu2019} or the LBM benchmark kernel suite~\cite{wittmann2018}.

Additionally, \walberla{} includes a rigid particle dynamics module to simulate particulate systems with the discrete element method (DEM) or hard-contact models.
This makes it comparable to particle frameworks like LIGGGHTS~\cite{LIGGGHTS}, GranOO~\cite{GranOO}, YADE~\cite{YADE}, PROJECTCHRONO~\cite{ProjectChrono}, and MercuryDPM~\cite{MercuryDPM}.

Similar to more general frameworks like AMROC~\cite{Deiterding2016,AMROC}, p4est~\cite{Burstedde2011} and Peano~\cite{Mehl2011}, however, \walberla{} has an explicit focus on HPC and also provides flexible adaptive mesh refinement and load balancing routines.

As a multiphysics framework, \walberla{} also enables efficient simulations of large-scale coupled fluid-particle systems in a monolithic way since both the fluid solver and particle modules are part of the same code and well interconnected.

The first prototype for \walberla{} was developed in 2007.
The framework has seen two architectural overhauls since, with the current iteration continuously developed since 2012 \cite{godenschwager_framework_2013} and released open-source as version 3.1 in 2017.
It is published under the GNU GPLv3 license~\cite{WalberlaGit} together with tutorials and documentation \cite{walberlaWebsite}.
Since the publication of Ref.~\citenum{godenschwager_framework_2013}, major new features like the rigid body dynamics module, adaptive mesh refinement, and code generation have been introduced as described below and are mostly included in version 4.2, the latest release. Version 5.0 is currently in development.

In this paper, we present the \walberla{} framework.
Its software architecture and components are briefly outlined and explained in \Cref{sec:overview}.
This is followed by a more detailed presentation of its individual components.
This includes the domain partitioning with the block forest in \Cref{sec:domain_partitioning}, and the discussion of data structures and kernels in \Cref{sec:block_data_and_kernels}.
In \Cref{sec:application}, the numerical methods and some respective applications are given.
\Cref{sec:codegen} explains the applied code generation techniques.
The supporting infrastructure of \walberla{} that assists developers and users is shown in \Cref{sec:infrastructure}.
Finally, in~\Cref{sec:extensions}, several projects that successfully build on and extend \walberla{}'s functionalities are highlighted.
\Cref{sec:conclusion} concludes the paper.

\section{Framework overview}
\label{sec:overview}

\begin{figure}[t]
	\centering
	\begin{tikzpicture}[scale=1,font=\sffamily]
	\baselineskip10pt
	
	\tikzstyle{moduleStyle}=[draw, minimum width=10.5cm, minimum height=1cm, rounded corners=0pt, thick, fill=black!10]
	\tikzstyle{submoduleStyle}=[draw, right=2mm, rounded corners=0pt, minimum height=6mm, fill=black!10]
	
	\node (m1) at (0, 0) [moduleStyle,fill=color1]{};
	\node [left=0.3mm of m1.east, align=right,white,font=\bfseries\sffamily]{core};
	\node (sm1) [right=0.3mm of m1.west, submoduleStyle] {MPI\vphantom{I}};
	\node (sm2) [right=0.3mm of sm1.east, submoduleStyle] {config\vphantom{I}};
	\node (sm3) [right=0.3mm of sm2.east, submoduleStyle] {timing\vphantom{I}};
	\node (sm4) [right=0.3mm of sm3.east, submoduleStyle] {math\vphantom{I}};
	\node (sm5) [right=0.3mm of sm4.east, submoduleStyle] {...\vphantom{I}};
	
	\node (m2) [below=1.75mm of m1, moduleStyle,fill=color2]{};
	\node [left=0.3mm of m2.east, align=right,white,font=\bfseries\sffamily]{block\\ forest};
	\node (sm1) [right=0.3mm of m2.west, submoduleStyle] {refinement\vphantom{Ig}};
	\node (sm2) [right=0.3mm of sm1.east, submoduleStyle] {load balancing\vphantom{Ig}};
	\node (sm3) [right=0.3mm of sm2.east, submoduleStyle] {buffer\vphantom{Ig}};
	\node (sm4) [right=0.3mm of sm3.east, submoduleStyle] {checkpointing\vphantom{I}};
	
	\node (m3) [below=1.75mm of m2, moduleStyle,fill=color3]{};
	\node [left=0.3mm of m3.east, align=right,white,font=\bfseries\sffamily]{block data\\ and kernels};
	\node (sm1) [right=0.3mm of m3.west, submoduleStyle] {fields\vphantom{Ig}};
	\node (sm2) [right=0.3mm of sm1.east, submoduleStyle] {GPU\vphantom{Ig}};
	\node (sm3) [right=0.3mm of sm2.east, submoduleStyle] {geometry\vphantom{Ig}};
	
	\node (m4) [below=1.75mm of m3, moduleStyle,fill=color4]{};
	\node [left=0.3mm of m4.east, align=right,white,font=\bfseries\sffamily]{numerical\\ methods};
	\node (sm1) [right=0.3mm of m4.west, submoduleStyle] {LBM\vphantom{Ig}};
	\node (sm2) [right=0.3mm of sm1.east, submoduleStyle] {RPD\vphantom{Ig}};
	\node (sm3) [right=0.3mm of sm2.east, submoduleStyle] {fluid-particle coupling\vphantom{Ig}};
	\node (sm4) [right=0.3mm of sm3.east, submoduleStyle] {PDE\vphantom{Ig}};
	
	\node (vm) [right=6.75mm of m1.east, fit=(m1.north east)(m4.south east), inner sep=0, moduleStyle, minimum 
	width=1cm,fill=color5] {};
	\node at (vm) [rotate=90,white,font=\bfseries\sffamily]{supporting infrastructure};
	
	\end{tikzpicture}
	\caption{Overview of the framework's software architecture and components.}
	\label{fig:framework_overview}
\end{figure}
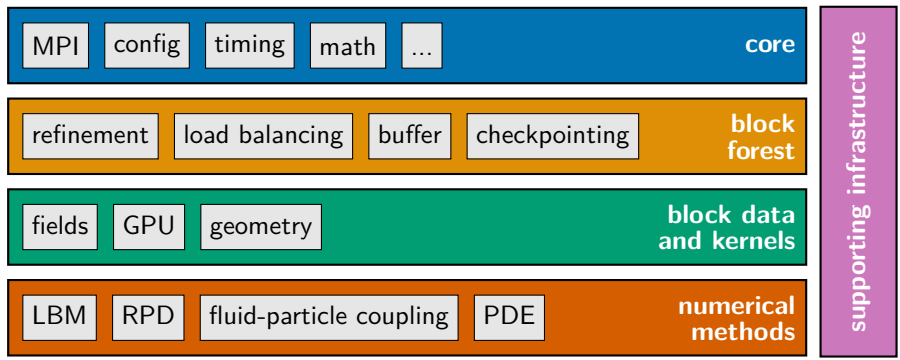

The \walberla{} framework is designed with a focus on modularity 
to enhance productivity, reusability, and maintainability.
An illustration of the software architecture is provided in~\Cref{fig:framework_overview}.
\walberla{} is implemented largely in C++(14) with Python extensions for enhanced usability.

The \textit{core} part contains the fundamental building blocks that are required for a parallel numerics framework.
This includes high-level abstractions around the MPI library and implementations of fundamental mathematical concepts, such as three-dimensional vectors or matrices.
Additionally, it contains functionality for handling configuration files that can be used to define simulation parameters in a convenient way.
Extensive time measurement options are available, providing the prerequisite for detailed performance testing and code optimizations.

The \textit{block forest} manages the domain partitioning into blocks, independent of data structures required for the actual simulation, and is described in~\Cref{sec:domain_partitioning} in more detail.
It provides functionality to statically or dynamically refine the partitioning, essentially splitting or merging blocks.
Determining optimal distributions of the blocks among the available processes in a parallel setup is a non-trivial task, and various load balancing techniques are available.
\walberla{} also provides components for checkpointing, i.e., to write and read a snapshot of a simulation such that it can be stopped and restarted or recovered after a hardware failure.
Finally, as information between the blocks has to be exchanged, versatile data buffers for sending and receiving data are implemented.

One major part of \walberla{} are kernels, i.e., algorithms that work on structured fields.
For this reason, much effort has been put into the careful design of efficient data structures and kernels, as is explained in detail in~\Cref{sec:block_data_and_kernels}.
This includes support for shared-memory parallelization and vectorization, as well as for specialized hardware like graphical processing units (GPUs).
For convenient handling of complex domains and boundary setups, we provide various functions to define geometries and to work with mesh files.
The most prominent example of a numerical method available in \walberla{} that makes extensive use of all those functionalities is the lattice Boltzmann method (LBM).
It is briefly explained in~\Cref{sec:lbm}, also stating our achievements in terms of performance and scaling on various supercomputers.

Moreover, \walberla{} also offers a highly efficient rigid particle dynamics (RPD) module that uses the same domain partitioning concepts as LBM, however, working on completely different data structures.
The available algorithms for particle simulations and the corresponding recent accomplishments are presented in~\Cref{sec:rpd}.

By coupling the LBM with the RPD module, multiphysics simulations can be realized on a large scale and without compromising on performance or scalability.
The available fluid-particle coupling strategies are highlighted in~\Cref{sec:coupling}, together with information on a number of physical problems that they have been applied to.

Code generation has been demonstrated to be a viable tool for rapid development of efficient compute kernels.
With these techniques, zero-cost abstractions can be realized. When using pure C++, optimizations and method description can oftentimes not be cleanly separated without sacrificing performance.
In~\Cref{sec:codegen}, we present our recently added code generation strategies for kernels that work on field and particle data structures.

All these parts are backed by an extensive testing environment that ensures ongoing correctness of the implementation, a well-structured collaborative development environment, and support of a wide range of compilers and software systems.
Furthermore, a graphical user interface (GUI) is available to facilitate rapid development of new functionality.
A powerful Python interface supports the user by providing efficient ways for pre- and postprocessing of the simulations.
A detailed description of these components is given in~\Cref{sec:infrastructure}.

Due to its flexible and generic software design, \walberla{} has been applied successfully in several projects as a potent basis for various extensions.
Those range from finite-difference phase-field simulations to finite-element earth-mantle dynamics simulations, and are briefly summarized in~\Cref{sec:extensions}.

\section{Block-structured domain partitioning}

\label{sec:domain_partitioning}

The core data structure of the \walberla{} framework is the \texttt{BlockForest}. 
It is responsible for the domain partitioning and acts as a container for all data needed by the simulation. 
In this section, we give an overview of its implementation and functionalities.
More information can be found in Refs.~\citenum{godenschwager_framework_2013,Schornbaum2016,Schornbaum2018}. 

\subsection{Overview}
\label{sec:BlockForest}

\begin{figure}[tb]
	\centering
	\begin{tikzpicture}[font=\sffamily]
	\fill[color1] (0,0) rectangle ++(2,2);
	\fill[color2] (0,2) rectangle ++(2,2);
	\fill[color3] (6,0) rectangle ++(2,2);
	\fill[color4] (6,2) rectangle ++(2,2);
	\fill[color1] (2,0) rectangle ++(1,1);
	\fill[color1] (2,1) rectangle ++(1,1);
	\fill[color1] (2,2) rectangle ++(1,1);
	\fill[color2] (2,3) rectangle ++(1,1);
	\fill[color2] (3,3) rectangle ++(1,1);
	\fill[color2] (4,3) rectangle ++(1,1);
	\fill[color4] (4,2) rectangle ++(1,1);
	\fill[color4] (5,2) rectangle ++(1,1);
	\fill[color4] (5,3) rectangle ++(1,1);
	\fill[color3] (4,0) rectangle ++(2,2);
	\fill[color1] (3,0) rectangle ++(0.5,0.5);
	\fill[color1] (3,0.5) rectangle ++(0.5,0.5);
	\fill[color1] (3.5,0) rectangle ++(0.5,0.5);
	\fill[color2] (3,2) rectangle ++(0.5,0.5);
	\fill[color2] (3,2.5) rectangle ++(0.5,0.5);
	\fill[color2] (3.5,2.5) rectangle ++(0.5,0.5);
	\fill[color3] (3,1) rectangle ++(0.5,0.5);
	\fill[color3] (3.5,1) rectangle ++(0.5,0.5);
	\fill[color3] (3.5,0.5) rectangle ++(0.5,0.5);
	\fill[color4] (3.5,1.5) rectangle ++(0.5,0.5);
	\fill[color4] (3,1.5) rectangle ++(0.5,0.5);
	\fill[color4] (3.5,2) rectangle ++(0.5,0.5);

	\draw[step=2,black,thick] (0,0) grid (8,4);
	\draw[step=1,black,thick] (2,0) grid (6,4);
	\draw[step=0.5,black,thick] (3,0) grid (4,3);
	
	\fill[color1] (9,3.25) rectangle ++(0.5,0.5);
	\node[right] at (9.5,3.5) {process $P_1$};
	
	\fill[color2] (9,2.25) rectangle ++(0.5,0.5);
	\node[right] at (9.5,2.5) {process $P_2$};
	
	\fill[color3] (9,1.25) rectangle ++(0.5,0.5);
	\node[right] at (9.5,1.5) {process $P_3$};
	
	\fill[color4] (9,0.25) rectangle ++(0.5,0.5);
	\node[right] at (9.5,0.5) {process $P_4$};
	
	\end{tikzpicture}
	\caption{Illustration of the domain partitioning in \walberla{}. The whole computational domain is subdivided into cuboidal subdomains (blocks) that can differ in size but have to maintain a 2:1 size ratio with direct neighbors. These blocks are then distributed among the available (here four) processes. }
	\label{fig:domain_part_blocks}
\end{figure}
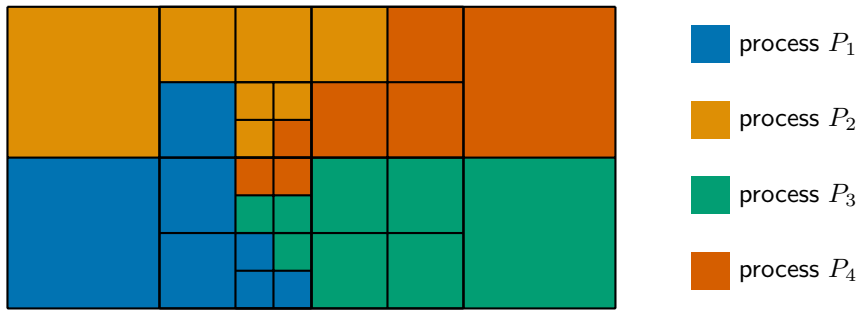

The starting point for the domain partitioning is a cuboidal simulation domain. 
This domain is partitioned in a regular fashion into equally sized subdomains. These intermediate subdomains, subsequently, act as root nodes for an octree, effectively forming a forest of octrees.
These individual octrees can be refined independently of each other as explained in~\Cref{sec:amr}.
The leaf nodes of the octrees make up the final subdomains of the simulation domain. 
The only restriction on the octree structure is a 2:1 size ratio, as illustrated in \Cref{fig:domain_part_blocks}, which is maintained between neighboring subdomains at all times.
Due to their shape, these subdomains are called \texttt{blocks} in the context of the \walberla{} framework.
In a parallel environment, every process is assigned one or more of these blocks. 
However, each block can only be assigned to exactly one process. 
The strict structure of the forest of octrees allows for an efficient distributed implementation which guarantees high scalability on large supercomputers. 
Every process only knows about blocks which got assigned to itself and has some information about neighboring blocks. 
All other blocks are unknown to a process. 
This particular detail ensures that the memory requirements needed per process to store the domain partitioning information does not increase with the number of processes~\cite{Schornbaum2016}.
However, the blocks are not only used to represent the domain partitioning, they are also data containers for arbitrary data. 
To use this storage, one has to implement the \texttt{BlockDataHandling} interface. 
This interface takes care of the creation, destruction, serialization, and deserialization of data allocated on every block. 
There are no restrictions on the data type and also no restrictions on the size, as long as it fits into the available memory. 
After registering the \texttt{BlockDataHandling} implementation with the \texttt{BlockForest}, the \texttt{BlockForest} creates an instance of the data for every block. 
This mechanism is used for all data required by the simulation, as described in \Cref{sec:block_data_and_kernels}.

\subsection{Adaptive refinement and load balancing}\label{sec:amr}

The capability to adaptively alter the domain partitioning is particularly useful for many engineering applications and thus a commonly sought feature in frameworks for numerical simulations~\cite{Dubey2014}.
When working on Eulerian grids, it allows maintaining a fine resolution wherever necessary, while at the same time coarsening the grid in less interesting regions, which often drastically reduces the computational cost and memory consumption of the simulation.
In \walberla{}, we fully support static and dynamic domain repartitioning.
The procedure that updates the \texttt{BlockForest} is carried out in four stages~\cite{Schornbaum2018}.
At first, blocks are marked for coarsening or refinement. 
This is done based on application-specific criteria that are registered as callback functions and therefore offer a maximum of flexibility.
Examples include the vorticity magnitude of the flow in fluid simulations and the number of particles in particle simulations.
In the second step, a proxy data structure is created from this refinement information. This data structure can be regarded as a lightweight version of the current \texttt{BlockForest}.
It only contains the new topological information without the memory-intensive simulation data and allows to specify the expected workload per block.
Based on this information, the proxy data structure can be load-balanced by redistributing the proxy blocks accordingly between all available processes.
For load balancing, commonly applied space-filling curves like Morton and Hilbert but also wrappers to libraries like ParMETIS~\cite{Schloegel2002,Parmetis} are available.
Due to the lightweight nature of the proxy blocks, even diffusive load balancing schemes, which require several internal iterations, can be carried out efficiently~\cite{Schornbaum2018}.
Finally, in the fourth step, the actual \texttt{BlockForest} is adapted according to the proxy data structure, which includes refinement and coarsening, as well as redistribution of the simulation data.

\subsection{Distributed memory parallelization}
\label{sec:mpi}
\begin{figure}	
	\centering
	\begin{tikzpicture}[font=\sffamily]

	\draw[fill=color2] (0,4.2) rectangle ++(10,2.5);
	\node[font=\bfseries\sffamily,color=white] at (5,6.2) {block-to-block};
	\node[](L1) at (2.5,5.6) {buffered schemes};
	\node[](L2) at (2.5,4.6) {\texttt{PackInfo}s};
	\draw[->] (L2) -- (L1);
	\node[](R1) at (7.5,5.6) {direct schemes};
	\node[](R2) at (7.5,4.6) {custom MPI data types};
	\draw[->] (R2) -- (R1);
	
	\draw[fill=color3] (0,1.1) rectangle ++(5,3);
	\node[font=\bfseries\sffamily,color=white] at (2.5,3.6) {buffered MPI};
	\node[right] at (0.3,3) {buffer system};
	\node[right] at (0.5,2.6) {\footnotesize efficient n-to-n communication};
	\node[right] at (0.5,2.2) {\footnotesize fixed \& variable message sizes};
	\node[right] at (0.3,1.5) {type-safe buffers};
	
	\draw[fill=color1] (0,0) rectangle ++(10,1);
	\node[font=\bfseries\sffamily,color=white] at (5,0.5) {MPI};
		
	\end{tikzpicture}
	\caption{\walberla{}'s distributed memory parallelization architecture.}
	\label{fig:comm_arch}
\end{figure}
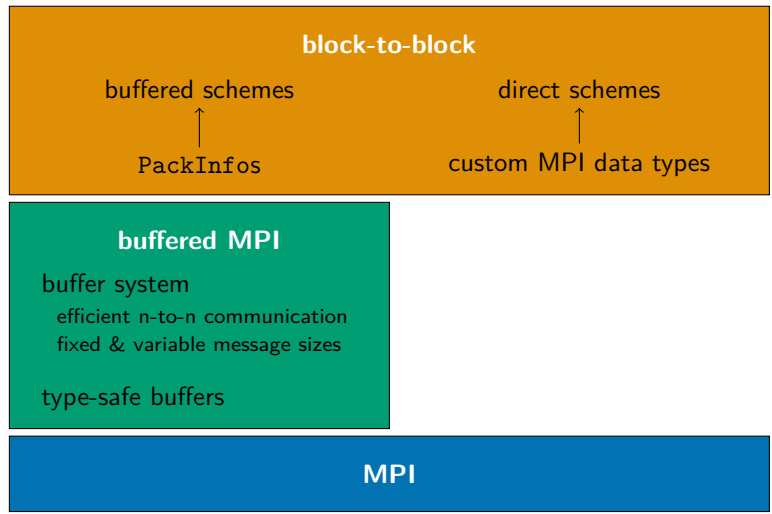

One central building block of the \walberla{} framework is its communication infrastructure. 
All of the distributed memory parallelization strategies are implemented on top of the Message Passing Interface~\cite{MPI} (MPI). The communication module is organized in a layered structure (cf.\ \Cref{fig:comm_arch}), going from low-level MPI functions up to classes that implement communication strategies. Only the topmost layer of the communication stack is specific to the block structure. The lower levels comprise a modular MPI serialization library that can be used independently of the \texttt{BlockForest}. Here, we describe the communication architecture from bottom to top.
At the lowest layer, \walberla{} provides thin abstractions on top of MPI. 
This layer is a C++ interface that integrates C++'s standard template library (STL) containers and basic \walberla{} data types like vectors and matrices with the MPI C interface. Additionally, it offers functions to broadcast input and geometry files to participating processes. Small files are read only once by the root process and are broadcast to all other processes to reduce the file system load. 

The ``buffered MPI'' layer offers flexible message serialization and deserialization capabilities.
In many communication scenarios, data has to be packed into a contiguous buffer from where it can be sent as a single message.
C++ provides a powerful input/output (I/O) stream library in its STL. 
Similarly, \walberla{} introduces buffer classes that can act as sources or sinks for serialization operations. By overloading the relevant operators, packing and unpacking is as simple for the user as interacting with the standard I/O streams. 
In this way, native C++ types, STL containers, and \walberla{}-specific types can be serialized. 
Error-prone, low-level data reinterpretations and address computations are encapsulated in this buffer layer and abstracted away from the user.
The send buffer internally holds a contiguous chunk of memory that can grow dynamically and can be transferred easily between processes via MPI functions. 
When a message is received, it is stored in a receive buffer object from where it can conveniently be deserialized. 
Our buffer abstractions enable us to transparently add metadata to the buffer.  This allows us to detect incorrect usage of the buffers. For instance, for type checking, the metadata specifies the data type of the buffer.
This metadata is then validated on the receiving side when the buffer is unpacked. 
Through this approach, type information is conserved even during the serialization and communication process. 
For release builds, the metadata is omitted to keep the messages as small as possible.

Also part of the ``buffered MPI'' layer is the \texttt{BufferSystem}.
It contains functionality for several forms of MPI point-to-point communication by keeping track of all communication partners of a process. It assumes that between these processes, a similar communication pattern is executed multiple times, which is, for example, the case for time-stepping algorithms.
The user specifies a set of processes to which messages should be sent to. For each communication partner, a buffer is allocated and reused in all the following communication steps. 
To receive MPI messages, the buffer system needs to know the set of processes from which messages are expected. 
The sizes of these messages may either be known beforehand by the receiver or only known to the sender of the message. The message size may either stay constant through multiple communication steps or vary over time. All cases are covered by the \texttt{BufferSystem} abstraction layer. 
Communication of messages with previously unknown size can be realized either by using \texttt{MPI\_IProbe} calls or by sending an additional message that contains the size of the following message(s). Depending on the cluster interconnect and MPI implementation, one or the other realization may give better performance.
A communication step is divided into several phases: First, for all expected messages, a non-blocking receive (\texttt{MPI\_IRecv}) is scheduled. This is done even before any messages are sent or packed such that the MPI system can already prepare for the expected messages, for example, by allocating buffers. The next step is to fill and send the outgoing buffers, again using non-blocking MPI functions. Note that the possibly time-consuming operation of serialization can be done after the receives have been scheduled. 
After the asynchronous sends and receives, computation on data that does not depend on the communication may be conducted. The last {\em wait} step finally receives and unpacks incoming messages. Internally, this phase is realized by calling a \texttt{MPI\_Waitany} function, that waits for any of the previously scheduled sends or receives to finish. If this function returns with a finished receive operation, the message is given to the user for deserialization while further messages can be received.
Usually, the received message can also be handled in a thread-parallel way. The user can register unpack functions (or functors) that are called asynchronously using OpenMP~\cite{OpenMP} parallelism as soon as a message arrives. The splitting into phases enables easy integration of the communication process into a task-parallel run time system such that computation and communication overlap.

Up to this stage, all abstractions are independent of \walberla{}'s block structure. 
The main interfaces the framework user interacts with are the block-to-block communication interfaces.
There are two approaches, one ``direct'' approach based on MPI data types and one ``buffered'' approach that builds on the ``buffered MPI'' subsystem.
We first describe the buffered part, shown on the top left in \Cref{fig:comm_arch}, where the data is serialized into buffers to reduce the number of MPI messages and thus costs due to network latency.
The user implements a \texttt{PackInfo} interface, where, given a local and a remote block, the user has to pack/unpack parts of data that is stored inside the block into buffers.
To illustrate this, consider the common case where each block stores parts of a large distributed array, a case described in more detail in \Cref{subsec:fields_sweeps}.
Using information about the neighboring block, the \texttt{PackInfo} decides which ghost region to pack or unpack. In adaptive mesh refinement (AMR) setups, the block also knows its own refinement level, such that grid transfer operations can be done during the packing/unpacking step. If the neighboring block stores a coarser grid, the data coarsening is already done on the sending side to reduce message sizes. 
Optionally, \texttt{PackInfo}s can implement a method to transfer data efficiently between two locally stored blocks without intermediate buffers. In scenarios where blocks differ in their computational load and \walberla{} balances the load by putting multiple blocks on a single process, this optimization is particularly useful.
\walberla{} provides \texttt{PackInfo}s for common block data, e.g., for the structured grid case described above, or the case of particle data. However, the user can store arbitrary data inside blocks. Additionally, by defining custom \texttt{PackInfo}s, the user can specify how these blocks are synchronized.

The top-most layer consists of {\em communication schemes} that encode the algorithmic structure of the communication. 
These objects use the layers below to provide a simple interface to conduct the communication either in a single call or use asynchronous constructs to hide communication behind computation. 
\texttt{PackInfo}s are registered with buffered schemes to define {\em what} is communicated, while the scheme defines {\em how} the communication is done. 

Besides the buffered communication schemes that pack data into buffers to reduce the number of MPI messages, a separate ``direct'' approach is available, shown on the right of \Cref{fig:comm_arch}. 
Instead of packing messages into buffers, the data to be communicated is described with custom MPI data types. 
For example, there are MPI data type definitions describing the array slices to realize ghost layer synchronization.
This MPI data type approach does not explicitly serialize and copy data and thus can be more efficient.
However, the buffered approach is easier to implement for custom block data and allows to send different data in a single message. For example, when simulating particulate flows, ghost layer information of a structured grid can be easily sent together with particle data. 
Therefore \walberla{} offers both options, letting the users choose whichever approach is best suited for their application.

\subsection{Checkpointing and resilience}

Systems with hundreds of thousands of compute cores are likely to fail from time
to time due to both hardware and software errors.
Especially future peta- and exascale systems are naturally expected to have
a lower mean time between failures due to their rapidly increasing size and complexity.
This trend can be observed for example when comparing the supercomputers Intrepid (installed 2008) and Sequoia (installed 2013), that were both initially ranked in the Top 5 of the TOP500~\cite{TOP500} list. 
Here, the reported mean time of interruption due to hardware failures decreased from 7.5 days  on Intrepid~\cite{Snir2014} to only 1.25 days on Sequoia~\cite{Dongarra2013}.
Applications in computational science and engineering typically require the vast amount of resources that only 
extremely parallel systems can offer.
With run times that are in the order of hours or even days, simulation software
is especially affected by the increased frequency of failing system components.
Restarting such simulations consumes an unnecessary amount of time and resources.
Therefore, we regard resilience as a central research topic to prepare for the next
generation of parallel computers. There are two categories of errors to be addressed:
soft errors --- such as bit flips that may remain unnoticed by the system --- and hard errors
like single node failures. Two approaches to handle such errors
at the software level are algorithm-based fault tolerance (ABFT)~\cite{Huang1984} and checkpoint-rollback recovery~\cite{Randell1975}.

ABFT aims to recompute lost data through carefully designed algorithms that are
in general tailored to the individual application. While such techniques have been successfully
employed in scalable applications \cite{Huber2016Resilience}, they are not easy to apply
in a general purpose simulation environment.

Checkpoint-rollback recovery is a straightforward and popular black-box solution to
recover from faults in simulation software \cite{zheng2012scalable,herault2015fault,Kohl_2019}.
During run time, the software regularly creates snapshots of the simulation data.
Upon failure, the application is restarted from the last available checkpoint and
the lost data is recomputed.
Checkpoint-based fault-tolerance techniques must be implemented carefully as the 
checkpoint-creation may suffer from bad performance due to the massive memory traffic
or slow disk I/O, as well as from massive memory requirements.
In Ref.~\citenum{Kohl_2019} we implement a distributed, in-memory checkpointing scheme
that is especially tailored to extreme-scale simulations like those that are performed with
the \walberla{} framework. We entirely refrain from writing to disk but distribute
the checkpoints in the main memory of the parallel processes.
We define small groups of processes, e.g., pairs of processes that store a snapshot
of their own simulation data and of the simulation data of the other group members.
This has two advantages: the memory footprint is only dependent of the size of the
process group and, therefore, constant and independent of the number of involved processes.
Second, the checkpoint can be created and loaded much faster in/from memory than on/from disk.
This technique guards the simulation very effectively against single-node failures
under the requirement that the process groups contain processes from different nodes.
We use an extension to the MPI standard called User-Level Failure Mitigation (ULFM)
\cite{Bland2013ULFM} that allows the user to detect and exclude failed MPI processes during run time and
to continue the running application.
In case of a process failure, all remaining processes load the last snapshot of their own subdomain and, if their
partner process failed, also the data of the failed process. Afterwards, the load balancer 
redistributes the data using the concepts described in \Cref{sec:amr}. Our implementation scales to parallel applications with more than $260\,000$ processes. 
In all scenarios, the checkpoint creation scales independently of the number of involved processes.

\section{Stencil codes on structured grids}
\label{sec:block_data_and_kernels}

As shown in~\Cref{sec:domain_partitioning}, \walberla{} provides a domain partitioning abstraction based on blocks.
In general, the framework user can store arbitrary data inside these blocks.
The framework additionally needs serialization information for this data in order to migrate blocks for load balancing. Also, the user has to define which parts of the data have to be communicated with neighboring blocks as ghost layers to arrive at a consistent state after the data has been updated locally.

There are two important distributed data structures that are built on this interface and already come with the framework: a distributed structured grid implementation and a distributed particle data structure for rigid body dynamics. In this section, we focus on the distributed grid infrastructure of \walberla{}, while the particle data structure is covered in~\Cref{sec:rpd}.

\subsection{Structured grids}

\begin{figure}
\centering
\includegraphics[width=\textwidth]{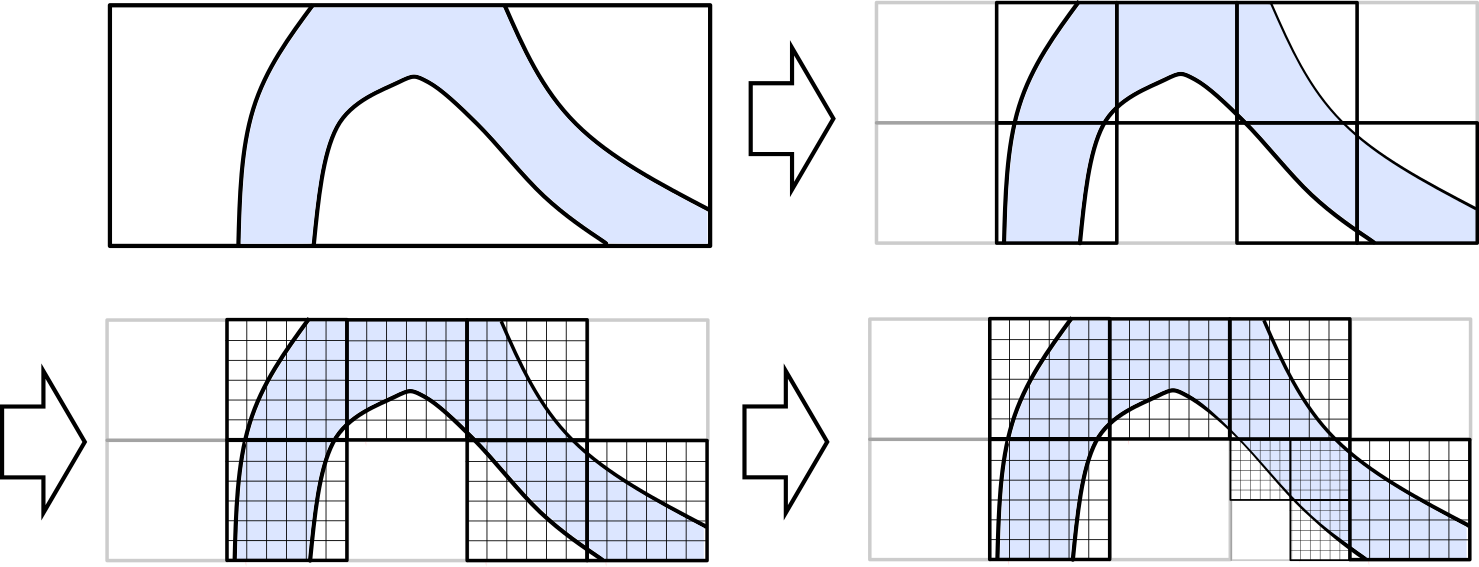}
\caption{Block-structured domain setup. From left to right: defining domain using a surface mesh, decomposition into
coarse blocks, allocation of cells in blocks, block refinement}
\label{fig:domain_setup}
\end{figure}

Originally, \walberla{} was designed to run stencil codes on structured grids, specifically simulations with the lattice Boltzmann method. For these applications, \walberla{} provides a distributed array implementation in which parts of the grid are stored locally in each block.
These blocks are synchronized using ghost layers.
Building on the generic \texttt{BlockForest} implementation that makes no assumptions about the data stored inside each block, \walberla{} also provides a specialized interface, called \texttt{StructuredBlockForest}.
This interface offers the user all necessary functions to translate block-local coordinates, i.e., indices in locally stored arrays, to global coordinates and vice versa.
The \texttt{StructuredBlockForest} assumes that each block itself is subdivided uniformly into grid cells and that every block has the same number of cells. These seemingly restrictive assumptions still allow for considerable geometric flexibility.
Consider, for example, the scenario shown in~\Cref{fig:domain_setup}. A lattice Boltzmann fluid simulation is set up in a domain defined by a surface mesh. First, the domain is partitioned into equally sized coarse blocks. Blocks that have no overlap with the domain are discarded. The blocks are then distributed to MPI processes during an initial load balancing step. After this, the actual simulation data is allocated inside each block. In this example, this is the field of probability density functions for the lattice Boltzmann simulation and a flag field marking fluid and boundary cells.
The grid can either be refined statically during simulation setup or dynamically, as is shown in the last step of \Cref{fig:domain_setup}. Refinement is done on the block level, meaning that every block still has the same number of grid cells.
The grid inside each block is extended by one or more ghost layers for distributed memory synchronization.
While the framework user could provide a custom grid implementation,
\walberla{} already comes with an array module, called \texttt{Field}, that provides a structured grid implementation together with all necessary integrations with other \walberla{} modules, most importantly, the \texttt{PackInfo} that links to the communication module, as described in \Cref{sec:mpi}.
This \texttt{PackInfo} supports the exchange of one or more ghost layers with neighboring blocks, as well as interpolation and restriction operations to handle neighboring blocks with different resolution.


\subsection{Fields and sweeps}
\label{subsec:fields_sweeps}

A central concept of \walberla{} is the separation of data and numerical kernels that operate on this data.
Kernels, called \texttt{Sweep}s in \walberla{}, are separate entities that can be optimized individually and adapted to special hardware architectures. Sweeps can also be selected based on properties of the processed block. It is possible, for example, to process some blocks on the GPU while others are handled by the CPU.
To implement a stencil algorithm on \walberla{}'s distributed grid, one only has to implement a serial or OpenMP-parallel numerical kernel operating on a \texttt{Field}.
A ghost layer exchange before or after the kernel call propagates the updated data to neighboring blocks.

The framework comes with several common sweeps that handle, for example, analysis and post-processing steps. Additionally, physics modules offer kernels for specific numerical methods as described in \Cref{sec:application}. Most prominently, the framework comes with highly optimized kernels for various lattice Boltzmann methods.

These kernels operate on a four dimensional array structure, the \texttt{Field}. Three dimensions are used to index the three spatial Cartesian coordinates, while the fourth dimension is used to store multiple values per cell, for example the components of a velocity vector or the distribution functions of a lattice Boltzmann grid.
The memory layout of a \texttt{Field} is configurable through the use of allocators.
The user can transparently switch between an array-of-structure (AoS) and a structure-of-arrays (SoA) layout.
The field layout is typically chosen to maximize the performance of the compute kernel. Thus, less time-critical parts of the framework such as set up and post-processing routines have to work with both memory layouts. Accessor functions hide this implementation detail and iteration constructs ensure that the field is traversed in linear order, independently of the chosen memory layout.
The iteration constructs provided by \walberla{} also allow for easy-to-use OpenMP-based shared-memory parallelization. The framework user can mark iterations over one or multiple fields for parallel execution. When OpenMP is activated, these loops are then processed in parallel.

To fully exploit single instruction, multiple data (SIMD) capabilities of modern CPUs, the memory layout has to be chosen accordingly and alignment restrictions have to be ensured.
Vectorized load and store operations benefit from, or even require, sufficient data alignment. For example, AVX vector instructions allow streaming (non-temporal) stores, which are crucial for the efficient implementation of lattice Boltzmann methods, but they can only be issued on aligned memory addresses.
At the end of each coordinate, a variable amount of extra padding elements is added, to ensure correct alignment of the following line.
\walberla{} comes with a SIMD module to vectorize numerical kernels manually. A thin abstraction layer on top of SIMD intrinsics for QPX, SSE, AVX and AVX2 allows the user to highly optimize a compute kernel with very close control over the hardware.

\subsection{GPU support}

Hardware accelerators, in particular, GPUs, are getting more and more important in HPC as some of the largest clusters in the world get the bulk of their compute power from GPUs.
Thus, it is crucial for an HPC framework to enable its users to port their simulations to GPUs without much effort.
\walberla{} uses the Nvidia CUDA toolkit, as currently all GPU-enabled supercomputing systems are based on Nvidia hardware.
\walberla{} comes with a GPU array implementation that closely mirrors its CPU counterpart, supporting different memory layouts and padding to ensure alignment.
Our implementation of GPU fields makes use of CUDA's support for two and three dimensional arrays, e.g., by using \texttt{cudaMalloc3D} and \texttt{cudaMemcpy3D}.
Choosing the same memory layout for CPU and GPU fields allows fast field transfers with a single \texttt{memcpy} call.
A typical GPU simulation in \walberla{} has mirror fields on CPU for each GPU field. These mirror fields are used for the initialization of the simulation data on CPU and are then transferred to the GPU.
The compute kernel is executed fully on the GPU, including boundary treatment. Only after simulating many time steps, the data is written back to the corresponding CPU fields, where post-processing steps are executed and result files are written out. With this approach, the geometry setup, post-processing, and I/O functionalities do not need to be ported to GPU.

GPU fields of \walberla{} come with an indexing abstraction layer. When manually developing a stencil kernel, the user has to decide how to map CUDA blocks and threads to array elements. A common approach is to map each thread to an array cell. CUDA threads are grouped into blocks. The block size influences the kernel performance in intricate ways, affecting register usage and occupancy. It is a tuning parameter that may be chosen differently on different GPU architectures.
\walberla{} comes with various indexing strategies that simplify this task.

The communication stack of \walberla{} is extended to handle GPU data as well.
When using CUDA-aware MPI implementations, data stored in GPU memory can be directly passed to MPI routines.
Thus, the ``direct'' communication approach based on MPI data types can be fully reused, as described in \Cref{sec:mpi}.
Since fields are mirrored to the GPU, the same MPI data type can be used to describe their ghost layers.
The ``buffered'' approach is adapted to support GPU buffers and their efficient transfer to pinned CPU memory.

\subsection{Geometry handling}
\label{sec:geometry_handling}


Handling parallel data structures poses problems to framework users that are not familiar with parallel computing. The complexity can be hidden from the user as long as only local operations on the data are required. The user writes local operations by defining sweeps over the full computational domain where values in a small neighborhood can be accessed. Ghost layer synchronization between sweeps ensures that each process has the necessary data of neighboring processes.
For evaluation, however, often non-local data accesses are required.
Consider, for example, the analysis of a velocity profile along a line through the domain. For convenience, \walberla{} provides functionalities to gather a user-defined subset of the simulation data on a single process. Although gathering of data should be avoided wherever possible, it can simplify the in-situ evaluation and steering of the simulation considerably.

For large scale simulations, I/O operations on simulation data can pose a significant bottleneck. While \walberla{} can read and write the full state of the simulation to disk efficiently with the help of MPI-I/O, the user can also post-process the data directly in the \walberla{} application to reduce output size.
For example, instead of dumping voxel data to disk, isosurfaces can be written out as a triangle surface mesh.
This technique is useful especially for multi-phase simulations where the interface between fluids is of interest. An integrated, custom marching cubes algorithm based on Ref.~\citenum{lorensen1987marching} builds meshes locally on each block. The algorithm also takes ghost layers into account, such that local meshes can be stitched together to a single mesh covering the full domain~\cite{bauer2015massively}. Triangles produced by the marching cubes algorithm have edge lengths in the order of the grid spacing, which is unnecessarily fine. They can be adaptively coarsened with an edge collapse simplification algorithm~\cite{garland1997surface}. A hierarchical tree-based gathering step collects local mesh pieces, runs the coarsening, and fuses them together to a single mesh.

To simplify the setup of the simulation and to define boundary conditions, \walberla{} comes with a {\em geometry} submodule. Boundary conditions are usually encoded in flag fields, where the cell type is encoded into bitmasks. Cells can be marked using geometric primitives like spheres, boxes, ellipsoids, or any geometric body obtained by boolean operators on these bodies. Additionally, 2D image files can be loaded and extruded to set up boundaries.
However, the most common and practical way to set up the simulation domain is through surface meshes.
To handle surface meshes, we use the OpenMesh library~\cite{OpenMesh}. It can load and store a variety of different polygonal mesh file formats. It is highly extensible, for example, by letting the user attach custom data to mesh elements.

Determining whether a block or lattice cell is covered by the mesh, is done by computing signed distances to the surface \cite{Jones1995, Berentzen2005}. \walberla{} contains an octree data structure to speed up distance computations of points to the surface mesh \cite{Payne1992}.
The surface mesh is used at various stages during a simulation setup.
First, it is used to set up the block structure by defining which blocks should be allocated and which regions should be refined (cf.\ \Cref{fig:domain_setup}).
Secondly, the mesh is used on the cell-level to define boundary conditions. Boundary type and boundary parameters, like velocity or pressure values, can be attached to vertices or faces of the mesh.

The setup routines store the geometry and boundary information in fields.
The straightforward way of handling boundaries is to iterate over all cells, extract the cell type from the flag field, and apply boundary treatment where necessary.
In the common case where the geometry is static and boundary treatment is only necessary for a small fraction of cells, a more efficient strategy is available.
A pre-processing step stores the indices of boundary cells in a list, optionally together with boundary information. For each boundary condition, a separate list is created such that each boundary can be treated by a separate kernel.

\section{Numerical methods and applications}
\label{sec:application}

To illustrate the flexibility and extensibility of the multiphysics framework, we show various problems that \walberla{} has been applied to.
We begin this section with a description of lattice Boltzmann simulations. 
This is the method \walberla{} was initially designed for, as is reflected in its acronym ``widely applicable lattice Boltzmann from Erlangen''.
Next, the methods contained inside the rigid particle dynamics module are presented, together with application examples.
Finally, the monolithic coupling of both, fluid and particles, for the simulation of particulate flows is discussed.

\subsection{Lattice Boltzmann method}

\label{sec:lbm}

We first briefly review the theory of the lattice Boltzmann method (LBM), then discuss performance aspects of the method, and report scaling results of lattice Boltzmann simulations.
Finally, we show some application examples where \walberla{} has been used successfully for the simulation of fluid flow.

\subsubsection{Overview}

The lattice Boltzmann method is a rather recent technique to simulate fluid flow based on a mesoscopic description.
A comprehensive introduction to the concepts of LBM can be found in Ref.~\citenum{Krueger2017}.
In LBM, the computational domain is discretized using a regular, evenly spaced grid described by a stencil as $DdQq$. 
This notation defines a $d$-dimensional domain where each cell contains $q$ probability density functions (PDFs) labeled $f_i(\mathbf{x}, t), i \in \{1, \dots, q \}$. 
The PDF $f_i$ represents the probability density of fluid particles moving from position $\mathbf{x}$ to $\mathbf{x}+\mathbf{c}_i$ with particle velocity $\mathbf{c}_i$.
Common stencil choices are $D2Q9$, $D3Q15$, $D3Q19$, and $D3Q27$, which are all supported by \walberla{}.
In the following description, we use lattice units where the cell positions $\mathbf{x}$ and the time step are integers.
In each lattice cell, macroscopic fluid properties like density and momentum density are obtained as moments of the PDFs:
\begin{equation}
\rho = \sum_i f_i,  \hspace{1cm} \rho \mathbf{u} = \sum_i \mathbf{c}_i f_i.
\end{equation}
The lattice Boltzmann update rule
\begin{equation}
f_i(\mathbf{x} + \mathbf{c}_i, t + 1) = f_i(\mathbf{x}, t) + \Omega_i \left(f_1(\mathbf{x}, t) , ..., f_q(\mathbf{x}, t) \right)
\end{equation}
evolves the system from time $t$ to $t+1$.
It consists of a collision step, described by a collision operator $\Omega_{i}$, and a subsequent streaming step where PDFs are propagated to neighboring cells.
Collision operators of methods considered here first transform the set of PDFs $\{f_i\}$ into a collision space. 
In this collision space, the components are relaxed towards equilibrium using convex combinations with rates that may vary for each of the $q$ PDFs. 
The most prominent collision space is the space of moments, yielding the formalism of multiple relaxation time (MRT) methods~\cite{dHumieres2002}. The resulting collision operator is
\begin{equation}
\label{eq:MRT}
\Omega_\text{MRT} \left[ \mathbf{f}  \right] = -\mathbf{M}^{-1} \mathbf{S} \left[\mathbf{M} \, \mathbf{f} - \mathbf{m}^{\text{(eq)}}  \right],
\end{equation}
where the transformation to moment space is represented by an invertible moment matrix $\mathbf{M}$.
The relaxation operation is expressed through a diagonal matrix $\mathbf{S}$. 
Special choices for $\mathbf{S}$ yield the widely known single-relaxation-time (SRT) and two-relaxation-time (TRT) methods~\cite{Ginzburg2008}.
The equilibrium distribution enters the equation through the vector of equilibrium moments $\mathbf{m}^{\text{(eq)}}$. 
\walberla{} contains highly optimized compute kernels for SRT, TRT, and various MRT methods. 
Additionally, cumulant methods are supported where cumulants instead of moments are relaxed to their respective equilibria~\cite{Geier2015}.

Another large class of lattice Boltzmann methods can be derived by taking one of the methods presented above and locally varying relaxation parameters based on some local fluid property like e.g., shear rates. This relaxation parameter adaptation is used to formulate turbulence models~\cite{Yu2005} or methods using entropic stabilization~\cite{KBC2015}. 
Several of these methods are available directly in \walberla{}. 
Since \walberla{} permits to vary relaxation rates locally by reading them from a separate field, methods of this type can be easily implemented and customized.

Furthermore, a large variety of boundary conditions for in- and outflow, pressure, velocity, and no-slip boundaries are available and can easily be used~\cite{Junk2008,Ginzburg2008}.
Alternatively, many applications can be more naturally represented by a periodic setup where the flow is driven by a body force.
In literature, various approaches to incorporate body forces into the simulation have been proposed~\cite{Guo2002,Krueger2017} and are implemented in \walberla{}. The user has the possibility to choose the most appropriate force model for the application at hand.

For many practical cases, employing a uniform grid throughout the whole computational domain can be prohibitive due to the huge amount of computational cost or memory that would be required.
In those cases, adaptive mesh refinement becomes a necessity.
For that reason, lattice Boltzmann methods for non-uniform grids have been developed where only in certain regions of interest the finest grid resolution is applied, and coarser grids are permitted elsewhere.
In \walberla{}, the method by Ref.~\citenum{Rohde2006} is implemented, which yields a mass-conserving scheme for local grid refinement.  

\subsubsection{Performance and scaling}
\label{sec:perf}
\begin{figure}[ht]
	\centering
	\begin{subfigure}[t]{0.48\textwidth}	
		\centering
		\includegraphics[width=\textwidth]{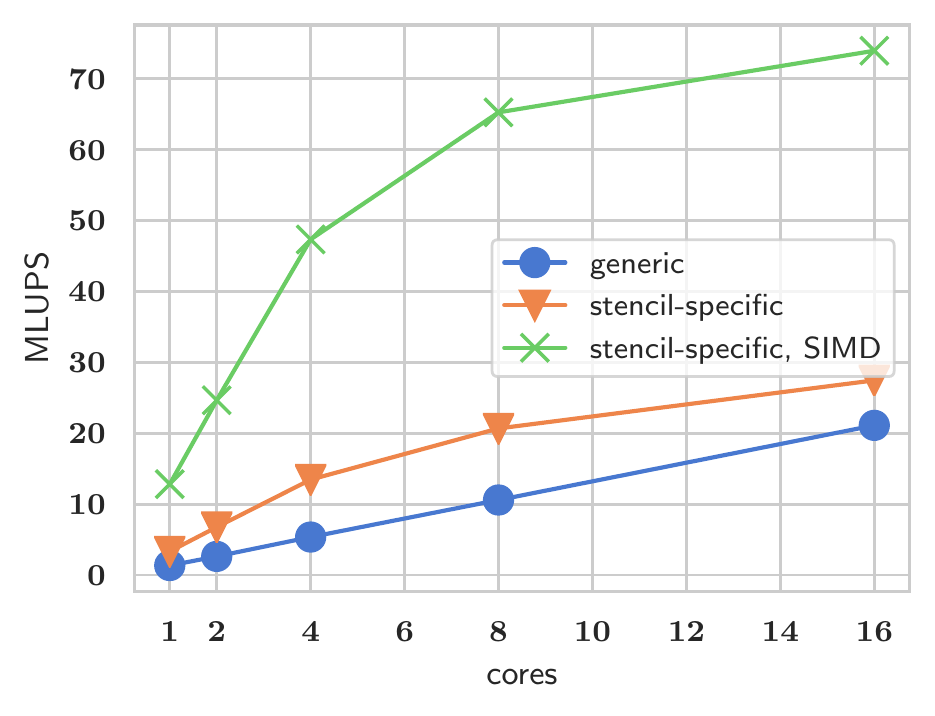}	
		\caption{Performance comparison of LBM kernels with different optimization levels on one node of JUQUEEN BlueGene/Q system.}
		\label{fig:plot_lbm_juqueen_singlenode}
	\end{subfigure}~\hfill
	\begin{subfigure}[t]{0.48\textwidth}	
		\centering
		\includegraphics[width=\textwidth]{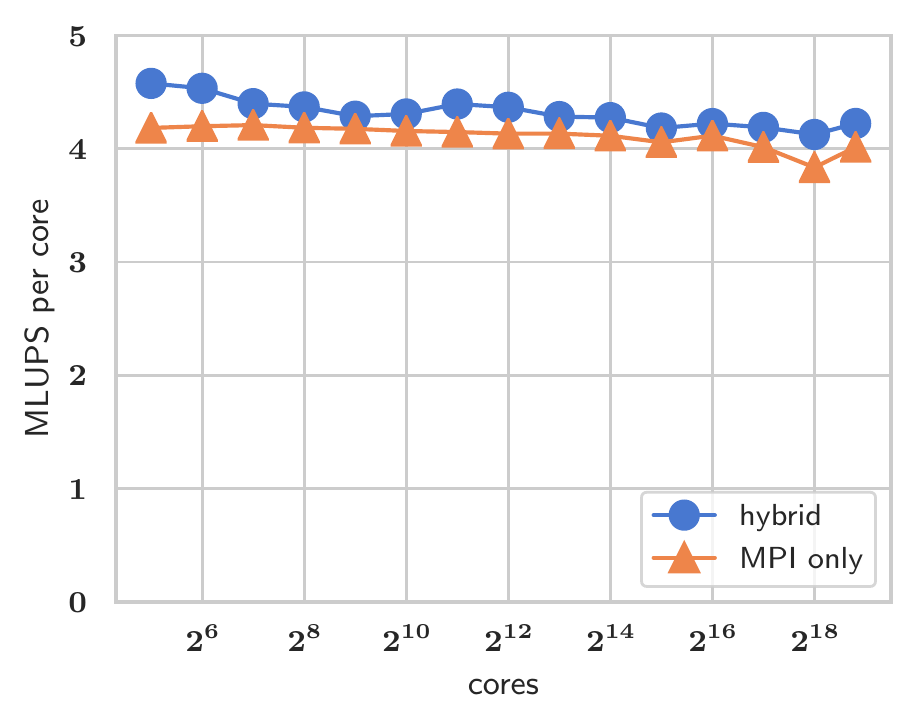}	
		\caption{Full machine weak scaling on JUQUEEN using pure MPI and hybrid (OpenMP+MPI) parallelization.}
		\label{fig:plot_lbm_juqueen_weakscaling}
	\end{subfigure}
	\caption{Performance and scaling experiments on JUQUEEN~\cite{godenschwager_framework_2013} using a $D3Q19$ TRT collision model. Performance is reported in million lattice updates per second (MLUPS).}
\end{figure}

To achieve high node-level performance, the LBM compute kernels make use of various node-level optimization techniques~\cite{zeiser2008introducing,donath2008performance, wellein2006single}.
As an HPC framework, \walberla{} aims to provide compute kernels that are highly optimized on the one hand, and on the other hand, are maintainable, flexible, and extensible. 
Using compile time polymorphism through template metaprogramming can be a reasonable trade-off between these somewhat conflicting goals. To get the best possible node-level performance, however, the compute kernel has to be specialized to a specific stencil and a set of manual optimization steps have to be applied.
The single node scaling results in \Cref{fig:plot_lbm_juqueen_singlenode} report the performance of a LBM $D3Q19$ TRT kernel on one node of the JUQUEEN supercomputing system. 
While the generic implementation, that uses static polymorphism to work for arbitrary LBM stencils, scales perfectly across one node, its overall performance is very low. 
The performance can be improved significantly by unrolling the loop that iterates over the distribution functions and thus specializing the implementation to a specific stencil. In this implementation, the stream and collision step is fused and the number of floating-point operations is reduced by manual common subexpression elimination. 
Explicit SIMD vectorization via intrinsics and further hardware-specific optimizations lead to the fastest kernel version that saturates the available memory bandwidth already at 8 from 16 available cores. For a detailed description of the optimization strategies and a performance model, we refer to Ref.~\citenum{godenschwager_framework_2013}.

In Germany, all current Tier-0 HPC systems --- SuperMUC-NG (Munich), Hazel Hen (Stuttgart) and JUWELS (Jülich) --- are supported by \walberla{}. The framework was also one of the first members of Jülich's High-Q club~\cite{HighQClub} of highest scaling codes on Jülich's supercomputing system. 
\Cref{fig:plot_lbm_juqueen_weakscaling} shows a weak scaling scenario on the JUQUEEN supercomputer to all 458\,752 cores of the full machine, using a pure MPI and a hybrid OpenMP+MPI parallelization approach. 
Perfect scalability is obtained even on these high core counts, due to \walberla{}'s fully distributed data structures and algorithms. 
\Cref{tbl:performance} shows that we support not only CPU-based clusters but can utilize GPU-based systems like the PizDaint supercomputer (Lugano, Switzerland) as well. 
\walberla{} utilizes advanced technologies, such as communication hiding with multiple CUDA streams and GPUDirect for asynchronous memory transfer from the GPU to the network adapter, to make full use of the capabilities available on PizDaint. 

\begin{table}[ht]
	\begin{tabular}{c|cccc}
		Supercomputer & Scale & Cells  & Performance  & Details \\
		&  & (billion) & (GLUPS) &  \\
		\hline
		JUQUEEN & 458\,752 cores & 1033 & 1937 & artery tree~\cite{godenschwager_framework_2013} \\
		JUQUEEN & 458\,752 cores & 886 & 890 & non-uniform grid~\cite{Schornbaum2016} \\
		SuperMUC & 32\,768 cores & 110 & 156 & non-uniform grid~\cite{Schornbaum2016} \\
		PizDaint & 2048 P100 GPUs & 34 & 2978 & uniform grid \\
	\end{tabular}
	\caption{Summary of scaling experiments of \walberla{} on various supercomputers for uniform and non-uniform grids.}
	\label{tbl:performance}
\end{table}

\subsubsection{Example applications}

\begin{figure}[ht]
	\centering
	\begin{subfigure}[b]{0.48\textwidth}	
		\centering
		\includegraphics[width=\textwidth]{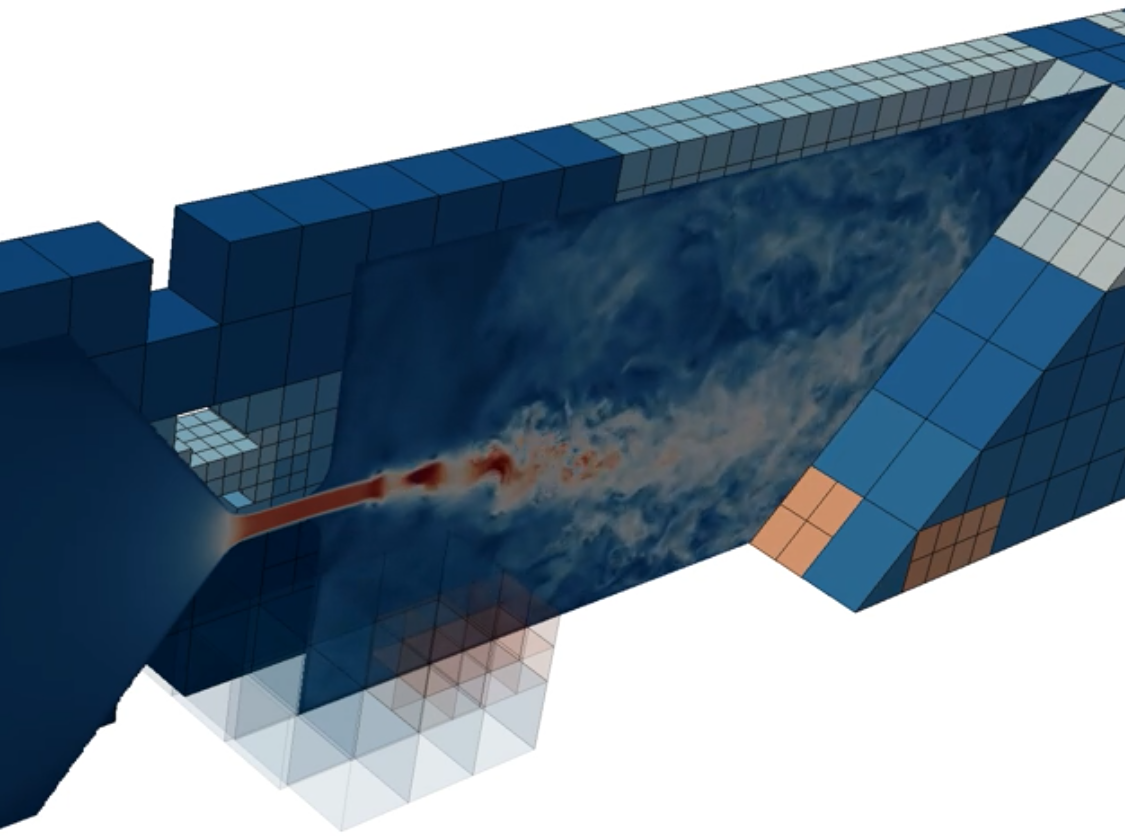}	
		\caption{Adaptive grid refinement for turbulent flow inside a vocal fold~\cite{Schornbaum2018}.}
		\label{fig:vocal_fold}
	\end{subfigure}~\hfill
\begin{subfigure}[b]{0.48\textwidth}	
		\centering
		\includegraphics[width=\textwidth]{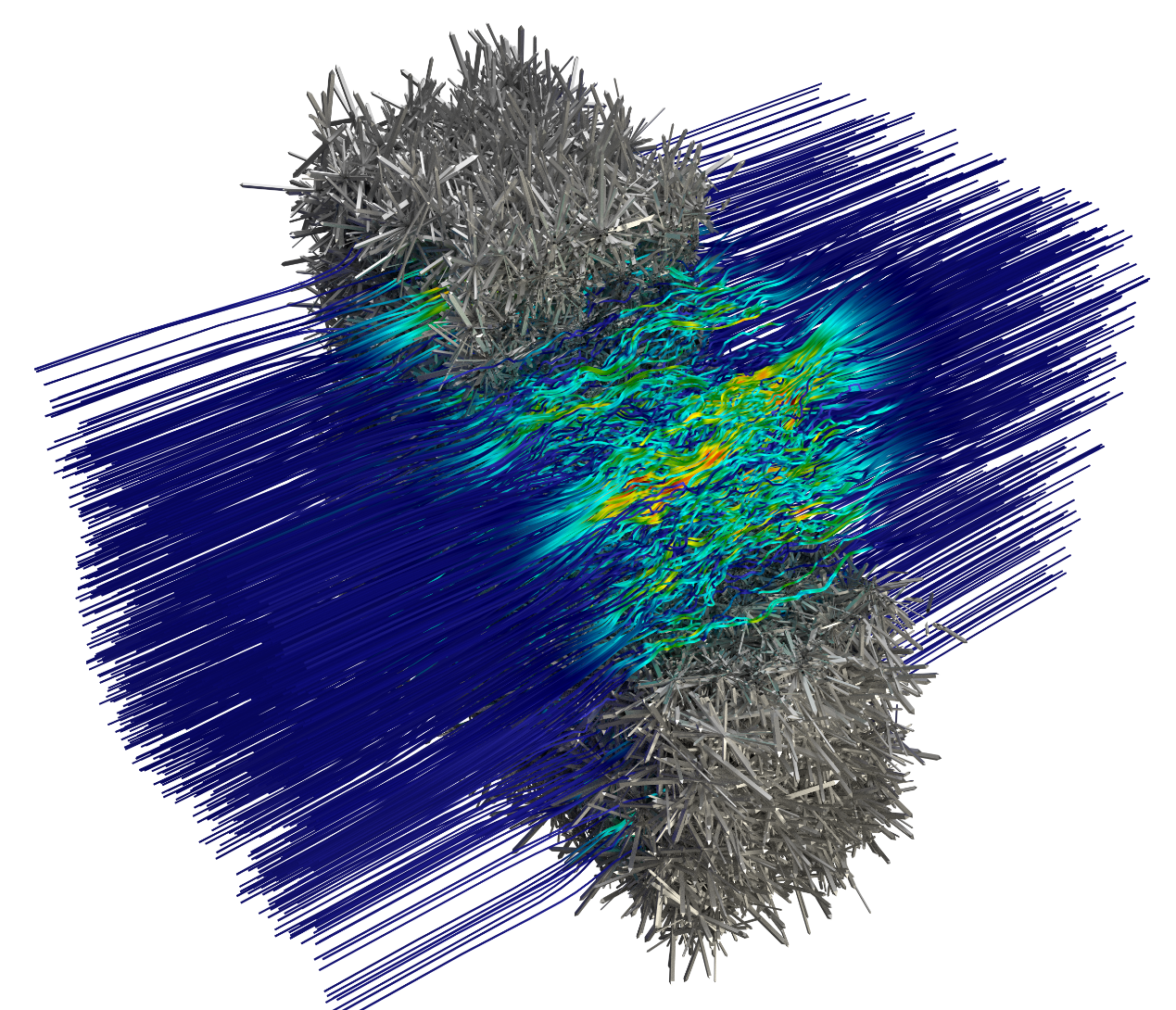}	
		\caption{Fluid flow through a porous crystal geometry~\cite{Gil2017}.}
		\label{fig:particulate_filter}
\end{subfigure}
\caption{Example applications of \walberla{}'s lattice Boltzmann solver for fluid simulations. }
\end{figure}

The \walberla{} framework has demonstrated its applicability to a wide range of fluid flow problems.
In Ref.~\citenum{godenschwager_framework_2013}, the flow through a coronary artery tree is computed on a massively parallel scale.
The difficulty of this setup arises due to the complex, sparse structure of the artery tree, covering only 0.3\% of the volume of the enclosing axis-aligned bounding box.
This leads to blocks that are only partially covered by the geometry and thus load balancing has been applied when setting up the simulation to reduce workload peaks.

Furthermore, the turbulent flow through a vocal fold, shown in~\Cref{fig:vocal_fold}, has been simulated to study the voice generation within the human throat.
In Ref.~\citenum{Schornbaum2016}, a static grid refinement approach inside the narrow channel of the fold has been applied to reduce the computational cost drastically.
This has been further improved in Ref.~\citenum{Schornbaum2018} by applying adaptive grid refinement techniques to maintain a fine resolution only in regions of turbulent vortices.

Since LBM is well-suited to simulate flow through complex geometries, it is commonly applied to study the flow through and above porous media.
This has been used in Refs.~\citenum{Fattahi_etal_2016,Fattahi_etal_2016b} and~\citenum{Rybak_2019} to derive new empirical models and interface conditions for laminar and turbulent flow in the vicinity of a porous bed, consisting of regularly or randomly arranged spheres. These arrangements are created with \walberla{}'s rigid particle dynamics module that is presented in \Cref{sec:rpd}.

The capability to generate boundaries from complicated meshes (cf.\ \Cref{sec:geometry_handling}) together with grid refinement has been used in Ref.~\citenum{Gil2017} to study a chaotic crystal structure that acts as a diesel particulate filter.
The resulting fluid flow and the porous structure are shown in~\Cref{fig:particulate_filter}.


\subsection{Rigid particle dynamics}
\label{sec:rpd}

The domain partitioning provided by the \walberla{} framework can also be used to implement a parallel rigid particle dynamics (RPD) simulation. Together the framework's math library, MPI wrapper, and load balancing capabilities provide a powerful foundation. In the following sections, the basic data structures and algorithms needed for particle simulations are introduced.

\subsubsection{Data structures}
The central piece of data for particle simulations is the particle data structure. All general information about particles such as position, rotation, and acting forces are collected in the \texttt{RigidBody} base class. Shape dependent properties like the radius for spheres and the edge lengths for boxes are added using inheritance from the \texttt{RigidBody} class. All particles are managed by the \texttt{BodyStorage} class that is responsible for the memory management of all particles. A \texttt{BodyStorageDataHandling} allows the user to add this new data type to the \texttt{BlockForest} (cf.\ \Cref{sec:BlockForest}). This way, all blocks automatically contain a \texttt{BodyStorage} data structure that stores all particles belonging to the subdomain that is covered by the corresponding block. Bodies with infinite extension, e.g., planes, are not stored on a block but in an additional global data structure. With this basic data structure in place, the simulation loop is able to update all particles continuously.

For parallel simulations, an approach similar to the ghost layer concept for grid-based simulations is used. All particles which belong to the subdomain of a block are stored on that block. These particles are called local particles. To enable parallel execution, one also needs to have information about particles that overlap the subdomain to identify possible interactions correctly. To achieve this, particles which belong to a different subdomain but overlap with another are copied to the overlapped subdomain. These particles are called ghost particles. Ghost particles cannot be manipulated directly as they are only copies of a local particle. Therefore, a synchronization algorithm is needed. The algorithm not only needs to keep all ghost particles synchronized with their corresponding local particle but also creates and destroys ghost particles as required. This synchronization algorithm is run at the end of every simulation time step.

Since the data structure needed for particle simulations, i.e., the \texttt{BodyStorage}, is implemented via the \texttt{BodyStorageDataHandling}, particle simulations with \walberla{} support all advanced capabilities based on serialization and deserialization of data. In particular, also load balancing can be used in the context of particle simulations. Therefore, the load balancing algorithms have been carefully adapted to particle simulations~\cite{Eibl2019}. run time load balancing can be done either based on the number of contacts per process or based on the number of particles per process. In load balancing algorithms that additionally take the load on the network into account, the number of ghost particles is used as an estimate for the communication costs.

\subsubsection{Algorithms}
A basic RPD simulation consists of four parts which are repeated in a simulation loop. In each loop iteration, the time is advanced by $\text{d}t$. The first part is the collision detection phase. The \walberla{} framework provides different algorithms for both the broad and narrow phase collision detection. For the broad phase, the user can choose between linked lists~\cite{Hockney1974,Allen2017} and hierarchical hash grids~\cite{Ericson2004,Erleben2005}. The narrow phase can be either done analytically or by using a combination of the GJK~\cite{Gilbert1988,Gilbert1990} and EPA~\cite{Bergen2003} algorithms. After successfully detecting all contacts, the interaction model tries to resolve these contacts. Soft contact models~\cite{CUNDALL1971,Cundall1979} and hard contact models~\cite{Preclik2015,Preclik2017} are available. After that, the simulation is integrated in time using Euler or Velocity Verlet integration schemes. For distributed memory parallel simulations, an additional synchronization algorithm has to be called. The framework provides efficient communication schemes for regular~\cite{Rapaport1991} and strongly polydisperse settings~\cite{Eibl2018}.

\subsubsection{Performance and scaling}
\begin{figure}[htbp]
  \centering

  \begin{minipage}[b]{.48\linewidth}
    \centering
    \includegraphics[width=\linewidth]{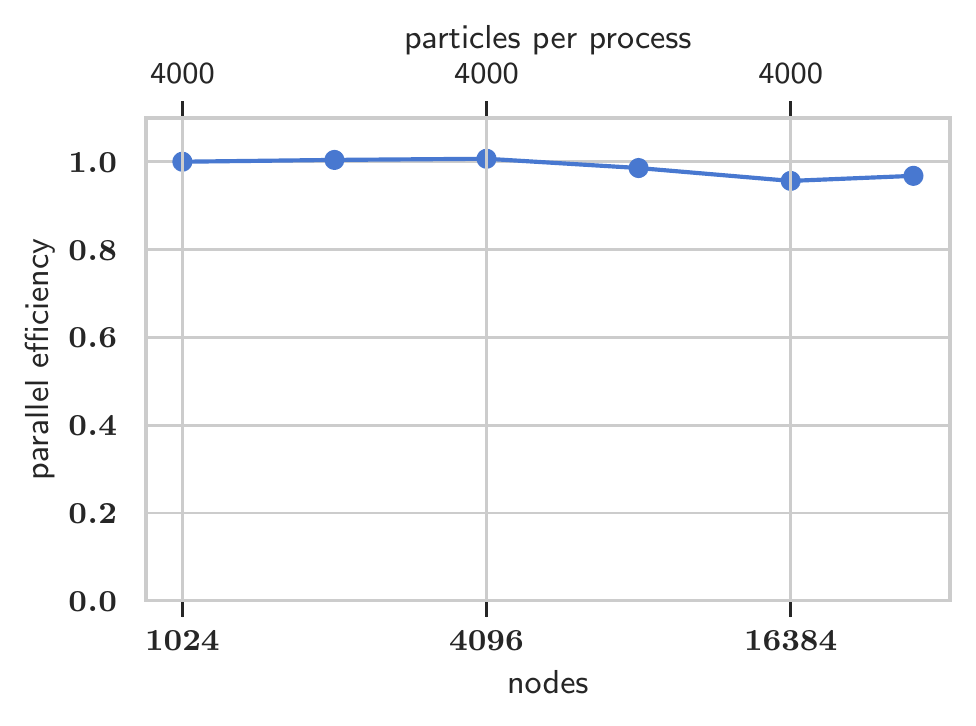}
    \subcaption{weak scaling}
  \end{minipage}
  \begin{minipage}[b]{.48\linewidth}
    \centering
    \includegraphics[width=\linewidth]{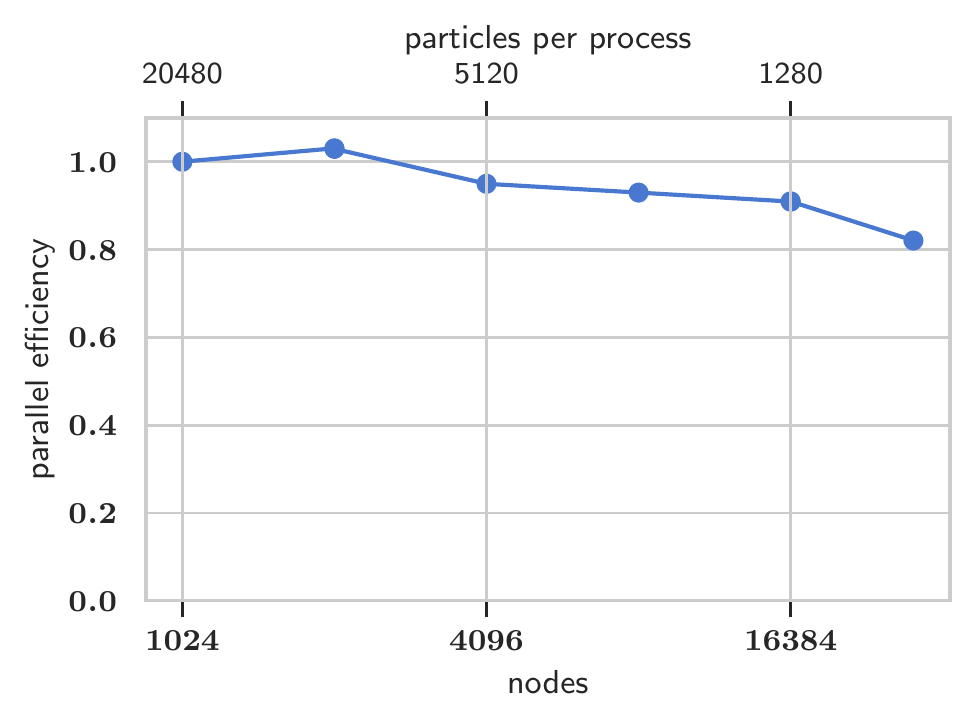}
    \subcaption{strong scaling}
  \end{minipage}

  \caption{Weak and strong scaling experiments for the rigid particle dynamics module. The simulation setup is an ensemble of particles that slide down a ramp.}
  \label{fig:pe_scaling}
\end{figure}

In the following, we will show the performance of the RPD module in a parallel environment by conducting weak and strong scaling experiments on the JUQUEEN supercomputer. This cluster is comprised of 16 cores per node with 4-way simultaneous multithreading (SMT). The full SMT capability is used to achieve maximal performance. The largest simulations carried out involve 458\,752 cores and 1\,835\,008 processes. For the simulation, spherical particles are generated on a hexagonal close packing lattice. The simulation domain is periodic in $x$- and $y$-direction. The $z$-direction is confined by solid walls. The gravity is applied under a \SI{30}{\degree} angle. This setup resembles a particle ensemble that is sliding down an inclined plate. The simulations are carried out with a different number of parallel processes, while the number of particles is scaled accordingly. \Cref{fig:pe_scaling} shows the results.

One can observe perfect weak scaling for up to 1\,835\,008 parallel processes which are the maximum number of processes available on that machine. The strong scaling is also carried out until the full machine is used. Good strong scalability is seen until 1280 particles per process. If the size of the simulation becomes too small, the communication overhead and the maintenance of all internal data structures exceeds the performance gain of adding additional cores. A detailed analysis of the scaling behavior can be found in Ref.~\citenum{Eibl2018}.

\begin{figure}[htbp]
  \centering
  \includegraphics[width=0.3\textwidth]{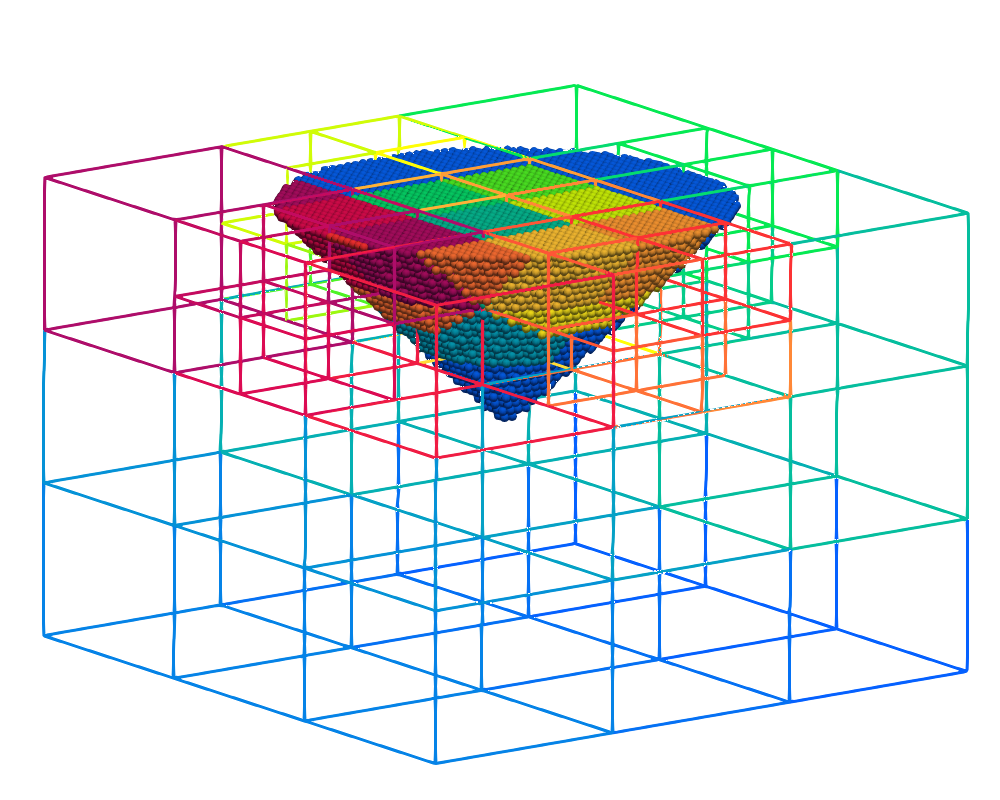}
  \includegraphics[width=0.3\textwidth]{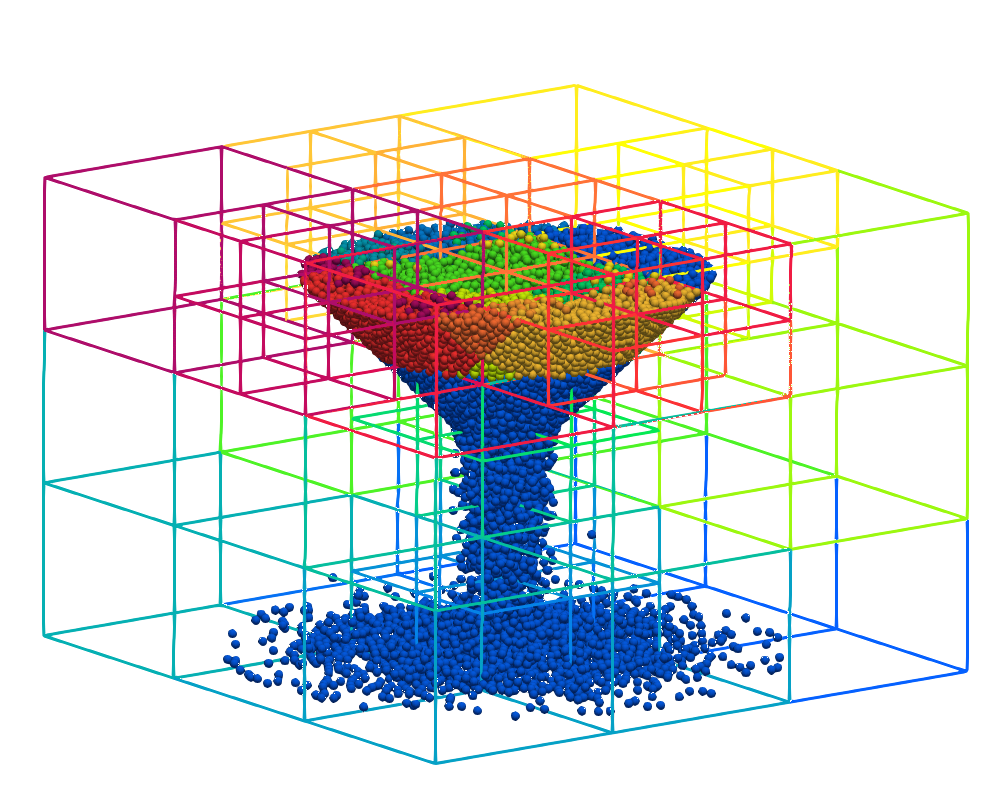}
  \includegraphics[width=0.3\textwidth]{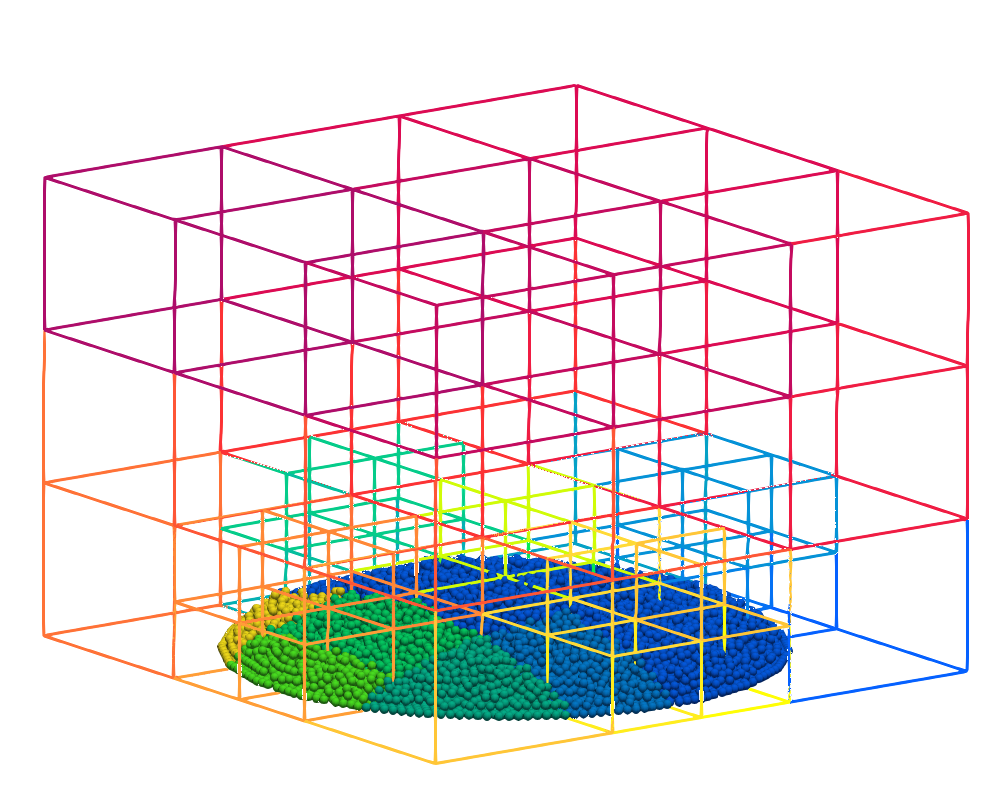}
  \caption{Hopper discharge simulation with 27 processes at three different points in time during the simulation. The subdomains are shown in wireframe. The subdomains as well as the particles are colored according to their process assignment.}
  \label{fig:pe_lb}
\end{figure}

The simulation of large particle ensembles often results in a very inhomogeneous distribution of particles over the simulation domain. Therefore, the framework also provides run time load balancing strategies to reassign subdomains to different processes dynamically. With this strategy, a more homogeneous workload over all processes is achieved. Only with run time load balancing enabled, highly inhomogeneous setups can be simulated efficiently. \Cref{fig:pe_lb} shows an example with a reduced number of particles for better visualization. Three stages of a hopper discharge simulation are shown with run time load balancing enabled. As the particles move from the upper part to the lower part, the domain partitioning has to continuously adapt to the simulation environment in order to guarantee an equal utilization of all available resources.

\begin{figure}[htbp]
  \centering
  \includegraphics[width=0.5\textwidth]{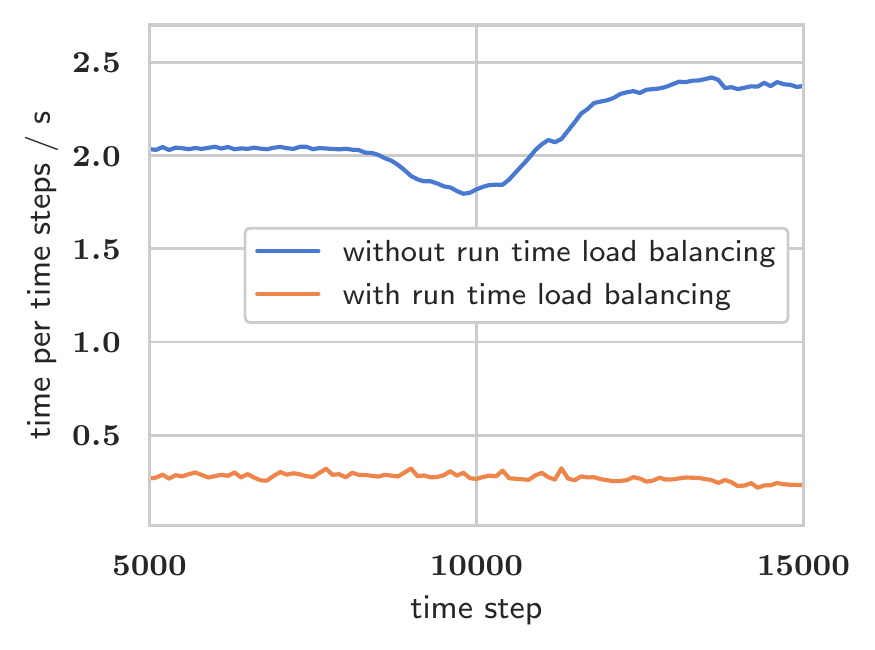}
  \caption{Hopper discharge simulation on the JUWELS supercomputer using 1296 cores. The computation time per time step without load balancing (blue) is considerably higher than for a simulation with run time load balancing enabled (orange).}
  \label{fig:pe_lb_runtime}
\end{figure}

For performance measurements, the hopper discharge simulation was run on the JUWELS supercomputer located in Jülich, Germany. The simulation size was increased to 8 million particles, and the simulation was run on 1296 processes in parallel. The time needed to advance the simulation by one time step is shown in \Cref{fig:pe_lb_runtime}. For the unbalanced simulation one can see that in the beginning, and in the end, the simulation performance decreases. In the middle, when half of the particles have fallen through the orifice, the simulation performance is best. However, by using load balancing, not only this deviation can be compensated but also the overall run time needed per time step can be drastically reduced by approximately a factor of 8. A detailed analysis of the load balancing capabilities of the \walberla{} framework for particle dynamics can be found in Ref.~\citenum{Eibl2019}.

\subsubsection{Applications}
The rigid particle dynamics module is used to gain insight into different physical processes. One area of research conducted with the \walberla{} framework is the analysis of random dense packings. With these simulations, information about the structural properties of granular materials can be extracted. Investigations of gravitational random dense packings of polydisperse spheres, together with validation against experimental results, are published in Ref.~\citenum{Schruff2016}.

By introducing attractive forces between particles, larger non-rigid particles can be composed out of smaller rigid particles. This approach has been used to study the behavior of carbon nanotubes under in-plane loading in Refs.~\citenum{Ostanin2018,Ostanin2019}.


\subsection{Fluid-particle coupling}

\label{sec:coupling}

By coupling our LBM fluid solver from \Cref{sec:lbm} to our particle simulation module from \Cref{sec:rpd}, we can efficiently simulate particle-laden flows at large scale.
The key aspect of this monolithic coupling is that we can directly access and update all underlying data structures since both solvers and the coupling are implemented inside the same framework.
This avoids possibly costly interfaces or abstraction layers, which would be needed if two separate frameworks had been coupled together.
In this section, we briefly outline the features of \walberla{}'s fluid-particle coupling and highlight several applications and extensions.

\subsubsection{Geometrically fully resolved simulations}\label{sec:mem}

\begin{figure}[htbp]
	\centering
	\includegraphics[width=0.5\textwidth]{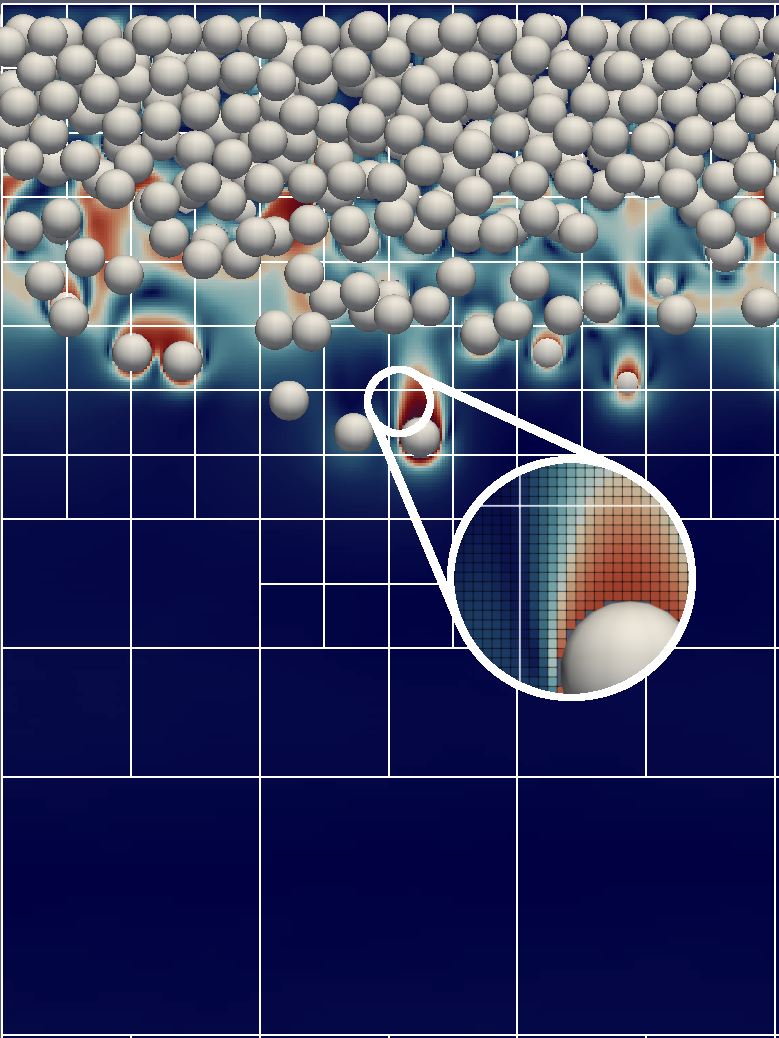}
	\caption{Simulation of the settling behavior of several geometrically resolved spheres with adaptive mesh refinement. The outline of the blocks, each containing $24^3$ cells, is shown with white lines. The inset also depicts the fluid cells around a sphere. The interaction between the particles is computed by our rigid particle dynamics module. The flow field, shown as a slice through the velocity field, is obtained via the LBM.}
	\label{fig:coupling_AMR}
\end{figure}

The interaction between a fluid and suspended particles is governed by the momentum exchange between these two components.
In literature, several modeling approaches exist that establish the coupling between fluid and particles.
Many interesting physical phenomena can only be correctly reproduced by coupling mechanisms where the geometric shape of the particles is fully resolved by the Eulerian fluid grid yielding highly accurate and thus predictive simulation results.
These so-called direct numerical simulation (DNS) approaches aim at representing the underlying physics from first principles.
In \walberla{}, we support the so-called momentum-exchange method of Refs.~\citenum{Ladd1994,Aidun1998}, and the partially saturated cells method from Ref.~\citenum{Noble1998}.
Both are well-established and commonly applied DNS coupling algorithms in the context of LBM.
These methods and their algorithmic details are thoroughly explained in Ref.~\citenum{Rettinger2017}, where a detailed comparison between them is presented.
Based on these results, we primarily employ the momentum-exchange method for geometrically resolved simulations of particulate flows and thus focus on this for the remainder of this section.
In this approach, the particles are explicitly mapped onto the Eulerian fluid grid by flagging cells inside the particle as \textit{solid}, whereas on \textit{fluid} cells the LBM is carried out.
The response of the particle to the fluid is established by computing the hydrodynamic interaction force acting on the individual particles via the momentum-exchange approach~\cite{Aidun1998}.
This force is then used in the rigid particle simulation module to update the particle position and velocity accordingly.
The other part of this two-way coupling, the response of the fluid to the particle's presence, is described by no-slip boundary conditions~\cite{Zou1997} along the surface of the particle, which are applied in the LBM fluid solver.
We can also use higher-order boundary conditions that increase the accuracy of the boundary treatment significantly~\cite{Rettinger2017}.
Due to the explicit mapping onto the grid, a re-initialization of newly uncovered cells is required where different techniques \cite{Aidun1998,Peng2016} are available in \walberla{}.
As shown in Fig.~\ref{fig:coupling_AMR}, it is also possible to use our fluid-particle coupling together with adaptively refined grids that maintain a high resolution around the particles, employing the AMR concepts explained in \Cref{sec:amr}.
Those regions that do not require a fine resolution are coarsened to increase the efficiency of the overall simulation.
However, the presence of particles on a block of the domain partitioning alters the block's resulting workload --- both in time and in space --- significantly.
Thus, to avoid unnecessary idle times of some MPI processes, we developed novel load balancing techniques to further increase the performance of these coupled simulations in Ref.~\citenum{Rettinger2019}.

\subsubsection{Simulated scenarios}

\begin{figure}[htbp]
\begin{subfigure}[t]{0.4\textwidth}
	\centering
	\includegraphics[height=4.5cm]{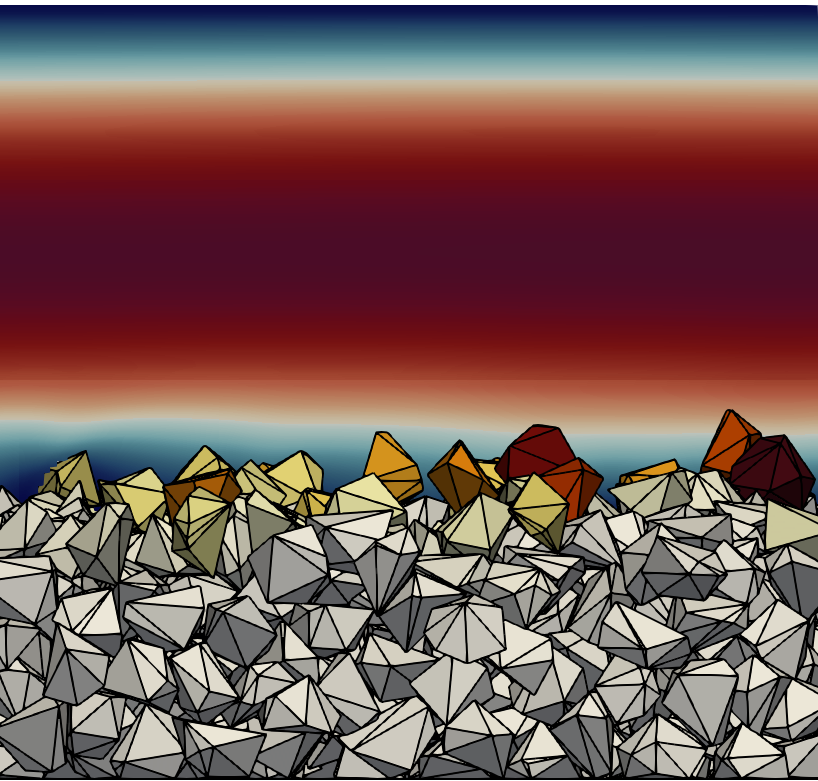}
	\caption{Simulation of particle dynamics in a riverbed. The particle shapes are randomly generated polygonal 
	meshes.}
	\label{fig:coupling_riverbed}
\end{subfigure}	
\hfill
\begin{subfigure}[t]{0.55\textwidth}
	\centering
	\includegraphics[height=4.5cm]{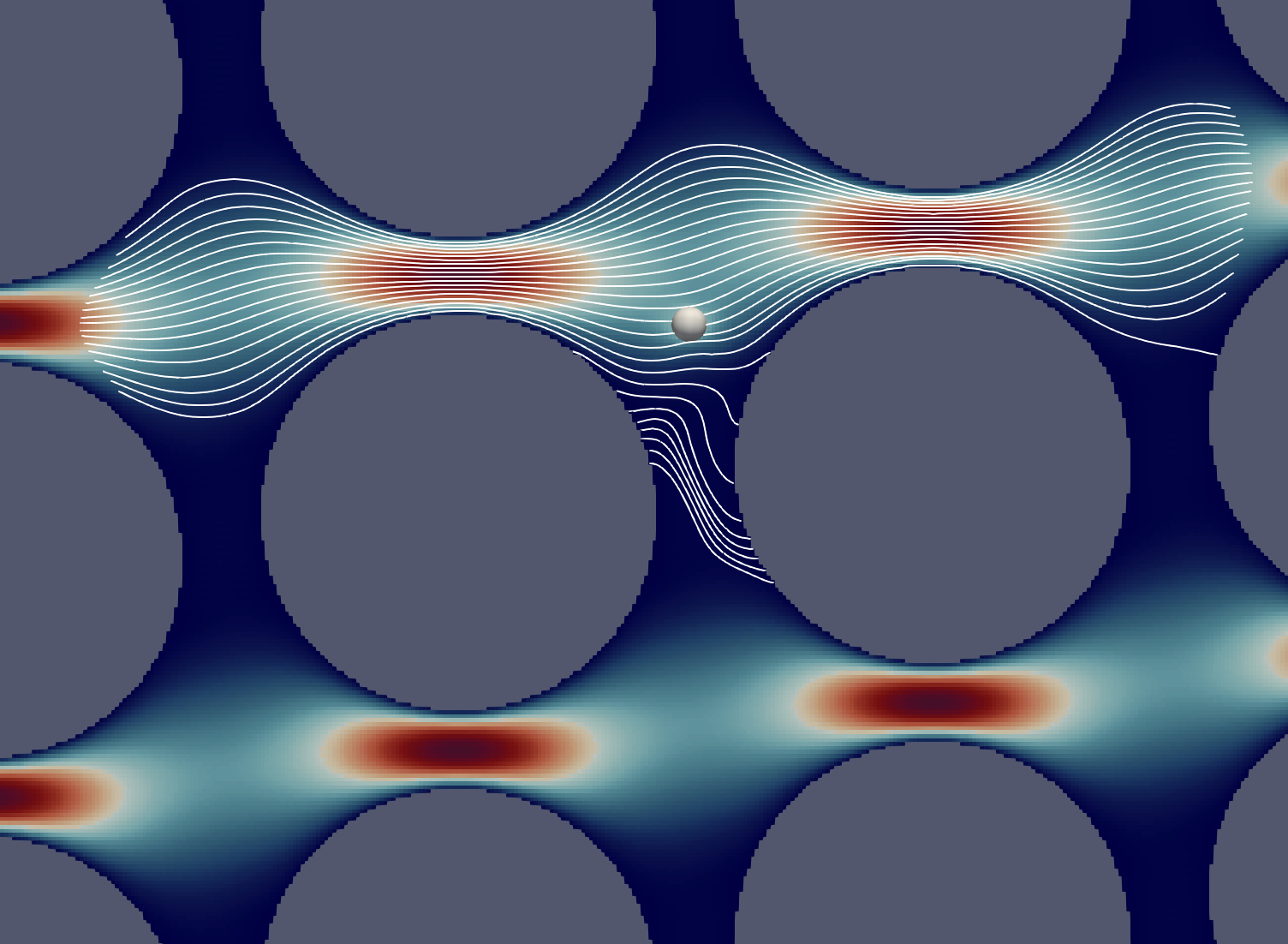}
	\caption{Simulation of a microfluidic particle-separation device. Higher velocity magnitudes are displayed in red, while lower 
	velocities are colored in blue. The streamlines are only visualized in the proximity of the moving particle.}
	\label{fig:microfluidic}
\end{subfigure}	
\caption{Examples of simulations with \walberla{}'s fluid-particle coupling.}
\end{figure}

The previously presented coupling functionalities, combined with our efficient fluid and particle simulation methods, allow us to investigate large particulate systems numerically on a massively parallel scale.
With such simulations, a more detailed insight into physical processes can be obtained than with classical laboratory experiments as information about forces on individual particles and the flow conditions around them are readily accessible.
In Ref.~\citenum{RettingerRiverbed2017}, we studied the formation of dunes in riverbeds.
As this necessitates a large number of particles, we applied static grid refinement above the bed to reduce the computational costs.
With this setup, we could show excellent scalability of our coupled simulations on the SuperMUC Phase2 supercomputer.
One of our simulations featured $16.2\cdot 10^9$ fluid cells and $7.7 \cdot 10^6$ spherical particles, making it one of the largest DNS of particulate riverbed flows carried out until now.
Since the applied coupling technique is also able to cope with different particle shapes, we are able to simulate particles described by mesh geometries.
This is shown in~\Cref{fig:coupling_riverbed}, where the sediment bed features particles of different shape given by randomly generated polygonal meshes.

The fluid-particle coupling implemented in \walberla{} is also utilized to simulate microfluidic particle-separation devices~\cite{Huang_2004} that are e.g., used in medical applications to separate red and white blood cells~\cite{McGrath2014}.
As shown in \Cref{fig:microfluidic}, a particle follows the fluid flow in between certain obstacles, such as pillars.
While these pillars are arranged equidistantly, they are laterally displaced perpendicular to the flow direction.
Due to this displacement, the fluid flows not only in horizontal direction, but also in vertical direction, as indicated by the streamlines in \Cref{fig:microfluidic}.
This effect can be used to separate particles based on their size.
While larger particles follow the flow horizontally, smaller particles follow the streamlines in vertical direction.
Our simulations allow us a detailed investigation of the underlying physical effects and the optimization of such particle-separation devices.

\walberla{} has also been employed to study the motion of microswimmers and their interaction with obstacles and with each other~\cite{Kuron_etal_2019a,Kuron_etal_2019b}.
Microswimmers~\cite{Elgeti2015}, also referred to as self-propelled particles or active matter, are micrometer-scale particles that propel themselves, converting fuel into directed motion.
Biological examples include microorganisms like bacteria and algae, while artificial realizations have, among others, been made from catalytic materials.
Microswimmers can interact with each other and with their surroundings not only hydrodynamically, but also electrostatically and phoretically. The latter term refers to interactions mediated by chemical concentration gradients, such as fuel depletion.
To incorporate the latter two interaction mechanisms, one can employ the method discussed for electrokinetic flows in \Cref{sec:ekin}.
If one is only interested in swimmers' hydrodynamic interactions, one can resort to effective propulsion models like the squirmer model \cite{Blake1971,Lighthill1952}, which has been implemented in \walberla{} using the momentum-exchange method of \Cref{sec:mem}.

\subsubsection{Volumetric coupling method}

These fully resolved simulations, although highly accurate, are computationally very expensive and are thus often not usable for simulations of particulate systems of industrial scale, like large fluidization reactors.
Therefore, we also developed and implemented algorithms that establish a coupling between the fluid and particles that are smaller than a fluid cell~\cite{Rettinger2018}.
For this unresolved coupling approach, empirical models for the fluid interaction like drag and lift forces have to be incorporated as they can only be obtained directly from a fluid simulation with sufficient resolution.
This reduces the required computational amount drastically and allows to simulate systems that are not accessible to fully resolved simulations.
A similar approach has also been used in Ref.~\citenum{Schruff2014} to study the infiltration of fine sediments into a porous sediment bed.
For that purpose, a deposition model has been implemented as the fine sediments gradually fill the holes inside the porous media.

\subsection{Coupling to non-hydrodynamic fields}
\label{sec:ekin}

Multiphysics applications typically concern themselves with more than just the hydrodynamic fields and suspended particles discussed up to this point.
\walberla{} is capable of solving partial differential equations (PDEs), like the Poisson equation for electrostatics, using standard iterative solvers like Jacobi, Gau\ss-Seidel, successive over-relaxation (SOR), or conjugate gradient (CG).
A spectral solver using the parallel fast Fourier transform library PFFT \cite{Pippig2013} is also available.
The iterative PDE solvers can also be combined to form a multigrid solver \cite{Bartuschat2015}.

\walberla{} has been used to simulate electrokinetic flows using the method of Capuani et al. \cite{Capuani2004,Rempfer2016}, which solves the diffusion of multiple solutes using a finite difference scheme, electrostatics by one of the solvers mentioned above, and hydrodynamics via LBM.
This method has been extended to incorporate particles via the momentum exchange method of \Cref{sec:mem} \cite{Kuron_etal_2016}.
The other fields typically cause a force acting on the fluid, while the opposite component of coupling is due to the advection of the solute with the fluid.


\section{Code generation} \label{sec:codegen}

There are multiple ways code generation can help in the development of a highly performant and portable code base. For example, as discussed in \Cref{sec:perf}, there is a tension between performance and maintainability. When developing high performance compute kernels, optimization oftentimes means specialization to a particular setup or hardware architecture. Highly optimized kernels are hard to read, extend, and maintain, while their performance is not portable. 
Traditionally, this problem has been tackled in \walberla{} by a strict separation of numerical kernels from the simulation data they operate on. The framework offers various kernel versions, fast specialized kernels for specific hardware and generic versions that are easy to extend and can be used for model development.
But still, each model has to be optimized separately for each target hardware. 
Consider, for example, the large number of different lattice Boltzmann variants, including different collision operators, stencils, force models, and storage patterns. Manually implementing and optimizing all variants leads to very large costs in terms of development time and effort. 
Thus, we aim to automate the process of kernel development and optimization through a code generation process. This process is described in \Cref{sec:pystencils}.

But not only kernels can be generated; data structures also offer a huge potential for code generation. If data structures are generated, they can be adapted to the current needs of the simulation, resulting in data structures as lightweight as possible. Besides, writing data structures by hand usually involves a huge amount of boilerplate code which is tedious and error-prone to write. With the newly designed \textit{Modular and Extensible Software Architecture for Particle Dynamcis (MESA-PD)}, we try to tackle this problem. A description of the concepts applied in this project is given in \Cref{sec:mesa_pd}.

\subsection{Pystencils}
\label{sec:pystencils}
The {\em pystencils} metaprogramming project~\cite{Pystencils} can generate highly efficient stencil codes for CPUs and GPUs based on a common high-level, symbolic description. On the symbolic level, the stencil formulation is derived from a continuous formulation of the PDE or the LBM collision operator.
This symbolic environment is based on the {\em SymPy} computer algebra system~\cite{SymPy} and allows the manipulation of the kernel by modifying expressions and assignments that are stored as syntax trees.
A set of common optimization techniques, like loop transformations and common subexpression elimination, are formulated as transformation on these trees. Backends then generate C/C++ code, CUDA code, OpenCL code or LLVM IR from this representation.
{\em pystencils} creates framework-agnostic code such that the kernel can easily be called from any environment that can interface with a C application binary interface (ABI). 
To simplify the interoperability with \walberla{}, the {\em pystencils\_walberla} package~\cite{PystencilsWalberla} offers a convenient integration of the code generation process into the C++ framework through the CMake build infrastructure.

Kernel code generation has already been applied successfully to \walberla{}-based LBM simulations and phase-field simulations of alloy solidification~\cite{bauer19phasefield} on CPUs and GPUs.

\subsection{MESA-PD}
\label{sec:mesa_pd}
The \textit{Modular and Extensible Software Architecture for Particle Dynamics (MESA-PD)} is an evolution of the current rigid particle dynamics module of the \walberla{} framework. It introduces a strict separation of data and algorithms and at the same time uses code generation techniques to automatically create parts of the software architecture. The main goals of the code generation are to free the programmer from writing most of the boiler plate code and to make all algorithms and data structures adaptable to the current simulation demands. 

The underlying technique of the MESA-PD project is the usage of Jinja~\cite{Jinja} templates to produce the C++ source files of the framework. First, the raw C++ source files are annotated with Jinja template instructions. And second, a high-level Python library is written to collect information about the simulation in a human readable way. This information is then processed and passed to the Jinja template engine, which uses this information to generate the final C++ source files. Although this approach is used throughout multiple places in the software framework, the most impactful place is the generation of the particle data structure. For this data structure, the Python library collects all properties the particle should have for the simulation. This ranges from position over applied forces to the temperature the particle currently has. This information is collected on a per-simulation basis, so the set of particle properties can be different between different simulations. The properties are also annotated with their data type as well as whether this property should be synchronized in a distributed memory parallel simulation. From this information, the Python library together with the Jinja template engine produces a data structure exactly tailored for the current simulation. But not only the data structure is adjusted. Also, all algorithms which rely on the data structure like synchronization, VTK output, serialization, etc.\ are automatically rewritten to support this new property set. This approach results in an overall optimized simulation which leads to better performance. At the same time, the generated C++ code still remains readable. It is also easier for domain specialists to adjust the framework without deeper knowledge about the inner details of the framework. An extended introduction into the techniques used in this project together with some code examples can be found in Ref.~\citenum{Eibl2019a}.

\section{Supporting infrastructure}
\label{sec:infrastructure}
As \walberla{} is developed by many different contributors, it is crucial to support developers by providing a high quality development infrastructure, as is outlined in the first part of this section. Additionally, we present our graphical user interface that assists developers in working with \walberla's data structures. We conclude this section with a description of our Python interface that can be used to setup, steer, evaluate and visualize simulations with \walberla.

\subsection{Development infrastructure}
\label{sec:ci}
A fundamental basis for sustainable software development is the use of a version control system (VCS). In the case of \walberla{}, a central Git~\cite{Git} repository is used. It is hosted on our own GitLab~\cite{GitLab} instance and is mirrored to GitHub~\cite{GitHub}. Another important feature to maintain usability and to ensure correctness of a software framework is continuous, thoughtful, and automated testing. This process is referred to as continuous integration (CI) and can also be followed by continuous delivery or deployment (CD). In our case, the capabilities of GitLab are used for these mechanisms. Every time a change is checked into the VCS, a pipeline consisting of several jobs is started where each job compiles and tests the software with different hardware and software configurations. This means different build options for the \walberla{} framework, e.g.,  building in release or debug mode, are tested together with different compilers and versions. At the time of this writing, tests are executed for different versions of the Clang~\cite{Clang}, GCC~\cite{GCC}, Intel~\cite{Intel}, and Microsoft Visual C++~\cite{MSVC} compilers. Of these different setups, 39 are currently tested at each commit and 121 at the extended nightly builds. This ensures that \walberla{} is compatible with a wide range of software stacks and also utilizes the fact that different compilers are capable of identifying different potential problems, i.e., display different warnings. After the framework has been compiled successfully, a wide variety of tests are executed to assure correctness. These include unit tests of single functions, integration tests of some components and full-framework tests executing some of the applications presented in \Cref{sec:application}. Additional tasks are integrated into the pipeline like static code analysis with Clang-Tidy~\cite{ClangTidy} or automated tools to find untested parts of the code, i.e., to measure test coverage. Extensive code documentation using Doxygen~\cite{Doxygen} is also automatically created and deployed to our website. Finally, performance tests are executed on a specific benchmarking machine to retrieve reproducible results, which are visualized on a Grafana~\cite{Grafana} instance. This mechanism is useful to track the impact of performance optimization over time and to identify problems as they arise.

To handle the large amount of configurations used for building and testing, Docker~\cite{Docker} containers are utilized. Therefore each job of the CI pipeline is started in a newly created instance of a Docker container, which is essential for reproducible testing. For our Windows-based tests, we similarly set up virtual machines which are automatically started on the build servers as needed.

\subsection{Graphical user interface}

Scientific codes are often shipped without a graphical user interface (GUI).
This has several reasons. First, the user base is expected to be comfortable with writing text-based configuration and script files.
Secondly, the high development effort for graphical user interfaces is often not worth their benefits, since powerful visualization software like Paraview~\cite{paraview} and VisIt~\cite{visit} satisfies most of the requirements of a typical scientific simulation code.
While \walberla{} is usually used to write out simulation data in VTK format for these visualization tools, it additionally does offer a graphical user interface.
Its GUI is not written to assist post-processing or simulation setup, but mainly aims to provide visualization of internal data structures to assist a developer who writes a custom application.
Consider, for example, the implementation of a new lattice Boltzmann boundary condition.
There, a detailed view of the LBM probability density functions (PDF) near the boundary is extremely helpful.
Simulation data is shown in-situ, letting the user proceed through the simulation step by step. In this way, the state after each time step can be inspected.
Additionally, the developer can set so-called GUI breakpoints.
These are constructs that call the GUI event loop at arbitrary points in the code, such that the program state at arbitrary points in the control flow can be visualized.

\begin{figure}[h]
\includegraphics[width=\textwidth]{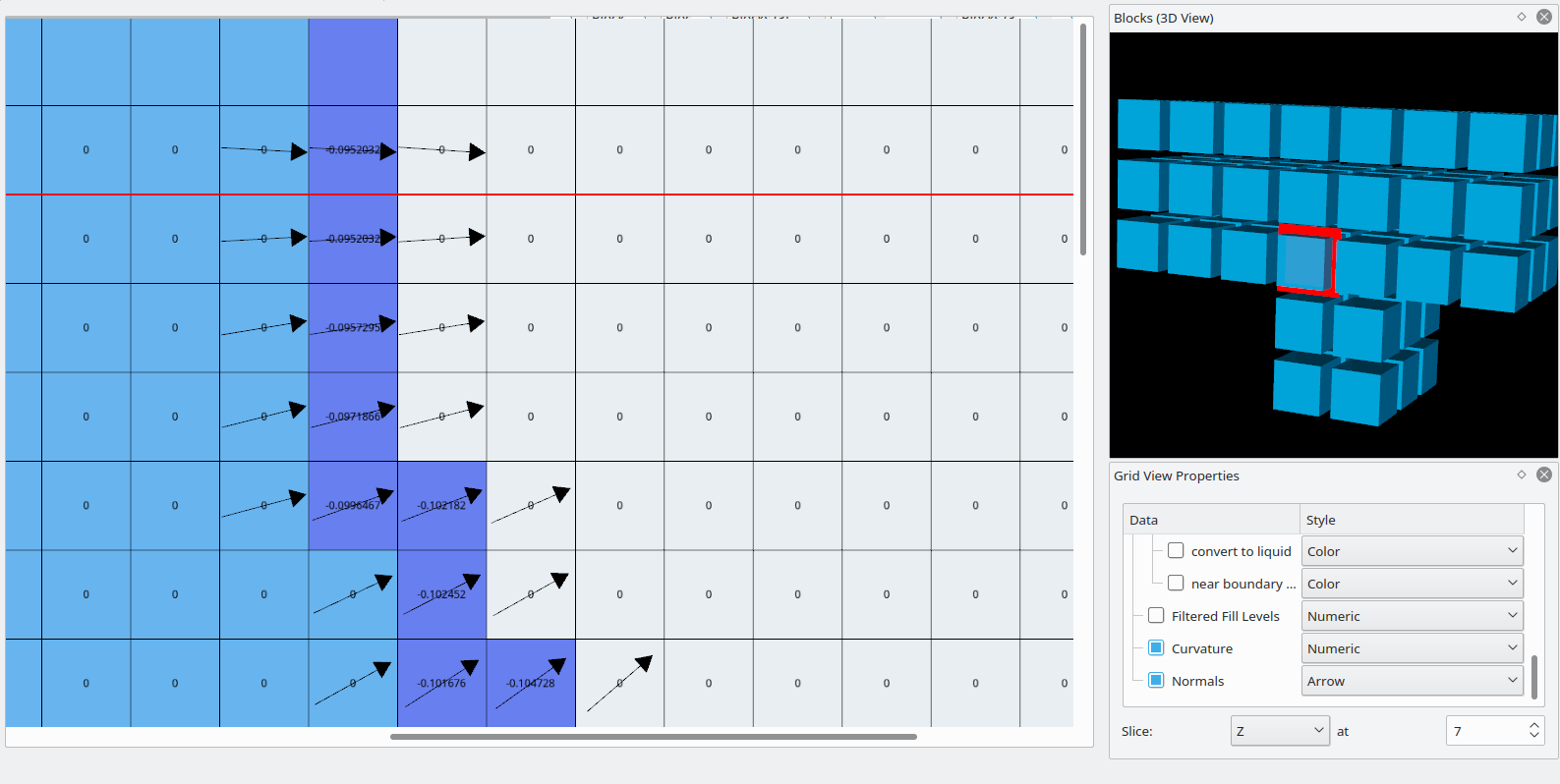}
\caption{Screenshot of \walberla{}'s graphical user interface. Exemplary display of a free surface LBM simulation, as described in \Cref{sec:fslbm}, showing field data like interface normals, curvature, and cell classification. Widget at the top right shows block partitioning, the data view on the left shows content of a single block.}
\label{fig:gui_overview}
\end{figure}

Since the types of data to be displayed varies from application to application, the GUI is implemented in a modular fashion.
One main widget displays the geometric layout of all allocated blocks in a 3D view.
All other widgets are specific to data types that may be stored in each block.
Each widget shows only the data of one user-selected block.
\walberla{} comes with a set of these specialized widgets, which can display commonly used data types such as scalar, vector, flag and PDF fields.
Additionally, the modular structure and the templates for structured and hierarchical data allow the developer to easily write new widgets for custom data if necessary.

\subsection{Python interface}

While the GUI's main goal is to assist the developer during debugging through visualization of internal data structures, it is not targeted at the user of the final simulation application.
Instead, \walberla{} provides a powerful Python interface for simulation setup, computational steering, as well as result evaluation and visualization~\cite{bauer2016python}.

\begin{figure}[h]
	\centering
    \includegraphics[height=6cm]{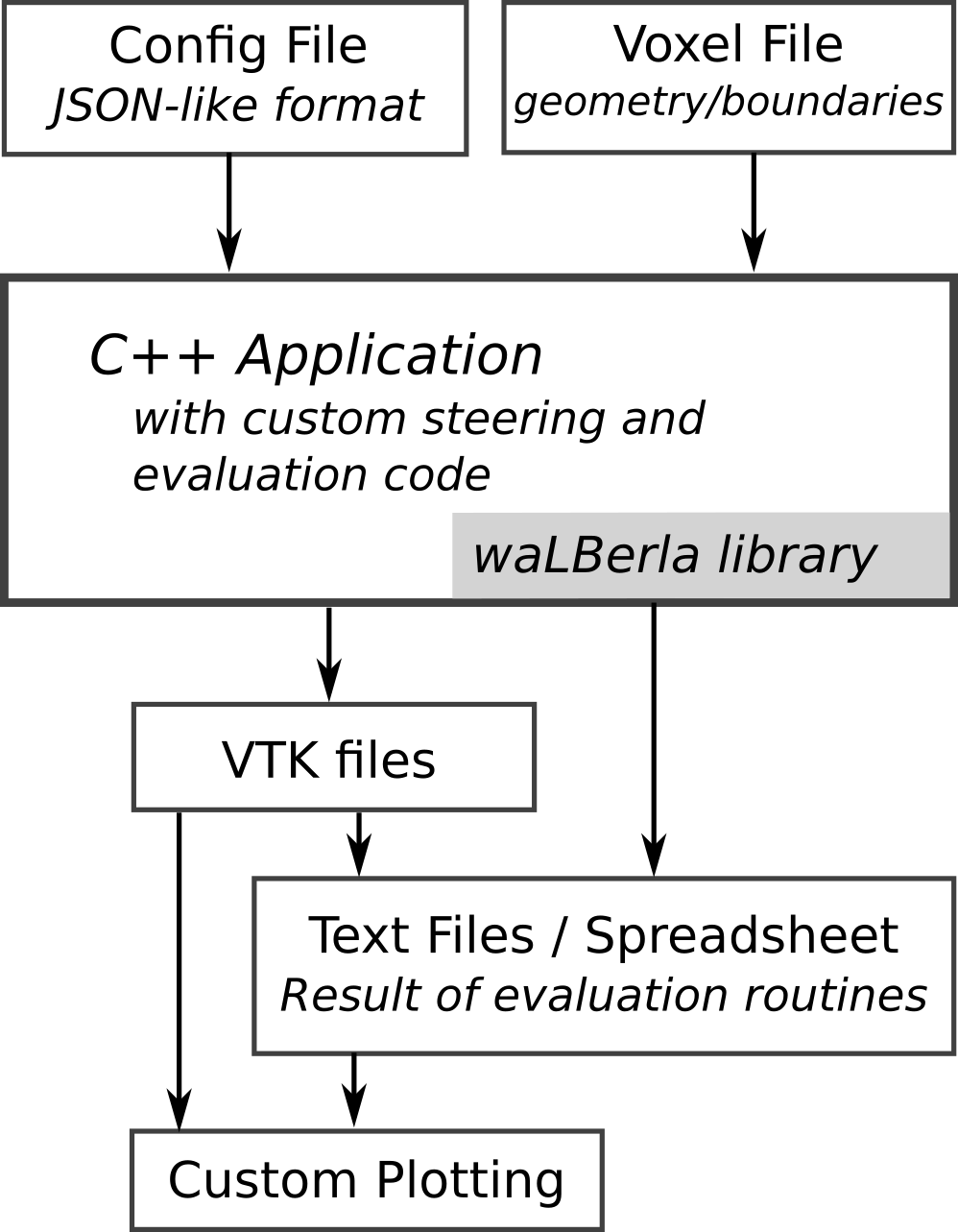}
    \hspace{1cm}
 	\includegraphics[height=4cm]{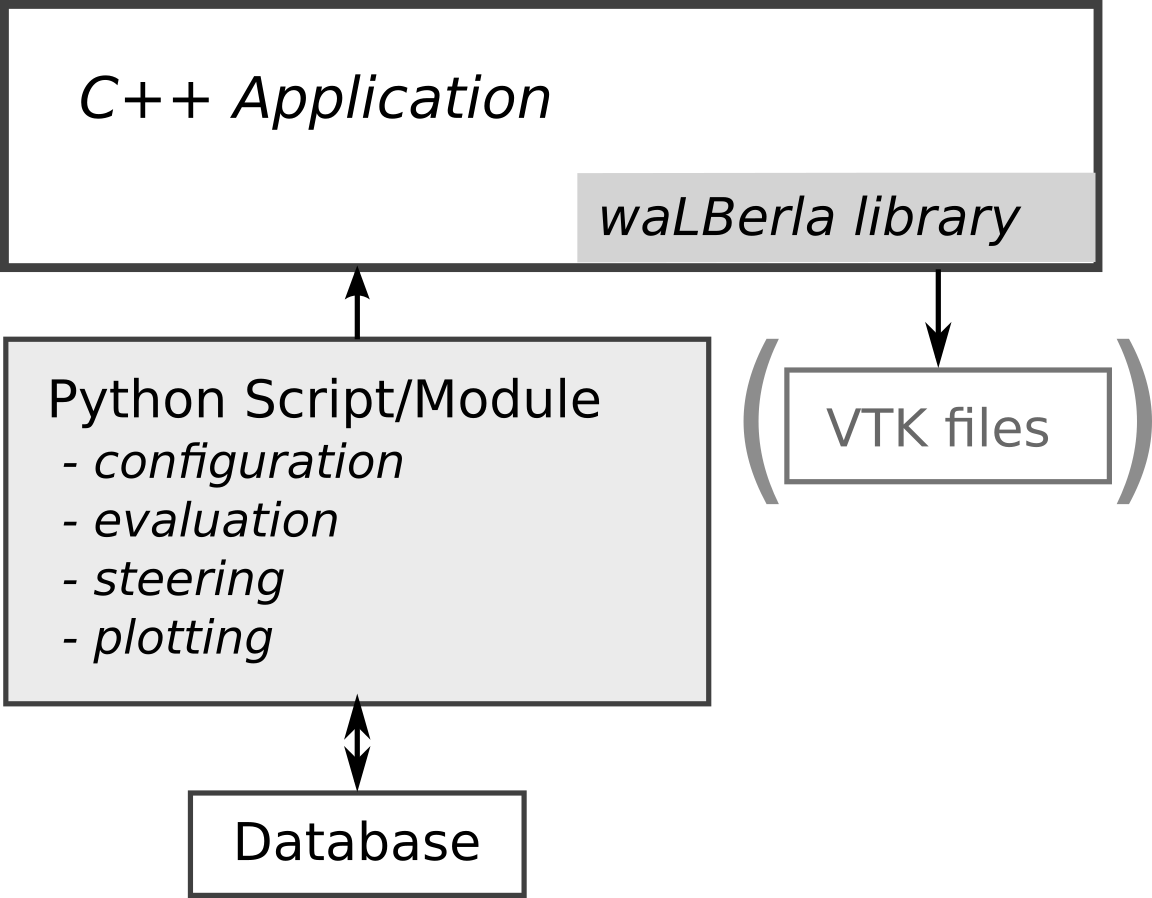}
     \caption{Simulation workflow without usage (left) and with usage (right) of Python Interface~\cite{bauer2016python}.}
     \label{fig:toolchain_python}
\end{figure}

Initially, the Python interface was intended as a replacement for the text-based configuration file.
In a pure text-based file format, there is no possibility to use mathematical expressions for parameter values, or define one parameter as a function of other parameters.
Switching from text-based to Python files as configuration files increases the expressiveness and flexibility for simulation setup drastically.
Technically, this is realized by embedding a Python interpreter into the \walberla{} application.
Additionally, we export central C++ data structures, like blocks and grids, to Python. This is done using the  {\em boost::python} library~\cite{BoostPython}. The locally-stored part of the grid can be accessed from Python as a NumPy \texttt{ndarray} \cite{NumPy}, making it compatible with the large scientific ecosystem of Python.
Thus, not only the setup process is simplified, but the entire simulation pipeline (cf.\ \Cref{fig:toolchain_python}).
Python callbacks can be registered to handle in-situ data evaluation tasks. At run time, the simulation data can be analyzed and results can be extracted. Instead of writing out large VTK files that contain the full simulation state, only relevant results are stored, for example in SQLite or non-relational databases that can be easily accessed from Python.
This approach extends the traditional C++ application that uses the \walberla{} framework, where the program flow is driven from C++.
Alternatively, \walberla{} can also be compiled into a Python module, such that a simulation application can be fully developed in Python.




\section{\walberla{} extensions}

\label{sec:extensions}

The \walberla{} framework has been used in several projects as a basis for massively parallel simulations.
In this section, we briefly outline some of these developed extensions that are not included in the official \walberla{} repository to highlight the flexibility and generality of our framework.

\subsection{HyTeG}
\label{sec:hyteg}
One software package that already utilized concepts and functionalities of \walberla{} is the Hybrid Tetrahedral Grids (\textsc{HyTeG}) framework~\cite{Kohl2019Hyteg} which aims at extreme scale simulations of Earth mantle convection. It uses the concept of strict separation between data structures and actual compute data and transfers this to unstructured triangle and tetrahedral meshes. On these so-called primitives, a hierarchy of structured fields is created which is well suited for geometric multigrid methods. In addition, parts of the \walberla{} codebase are used, like the CMake build system and a variety of functionalities from the core module. This includes the advanced buffer system explained in~\Cref{sec:mpi}, as well as basic math operations and the logging functionality.

\subsection{Free surface lattice Boltzmann method}
\label{sec:fslbm}
\begin{figure}[ht]
	\centering
	\begin{subfigure}[t]{0.48\textwidth}	
		\centering
		\includegraphics[width=\textwidth]{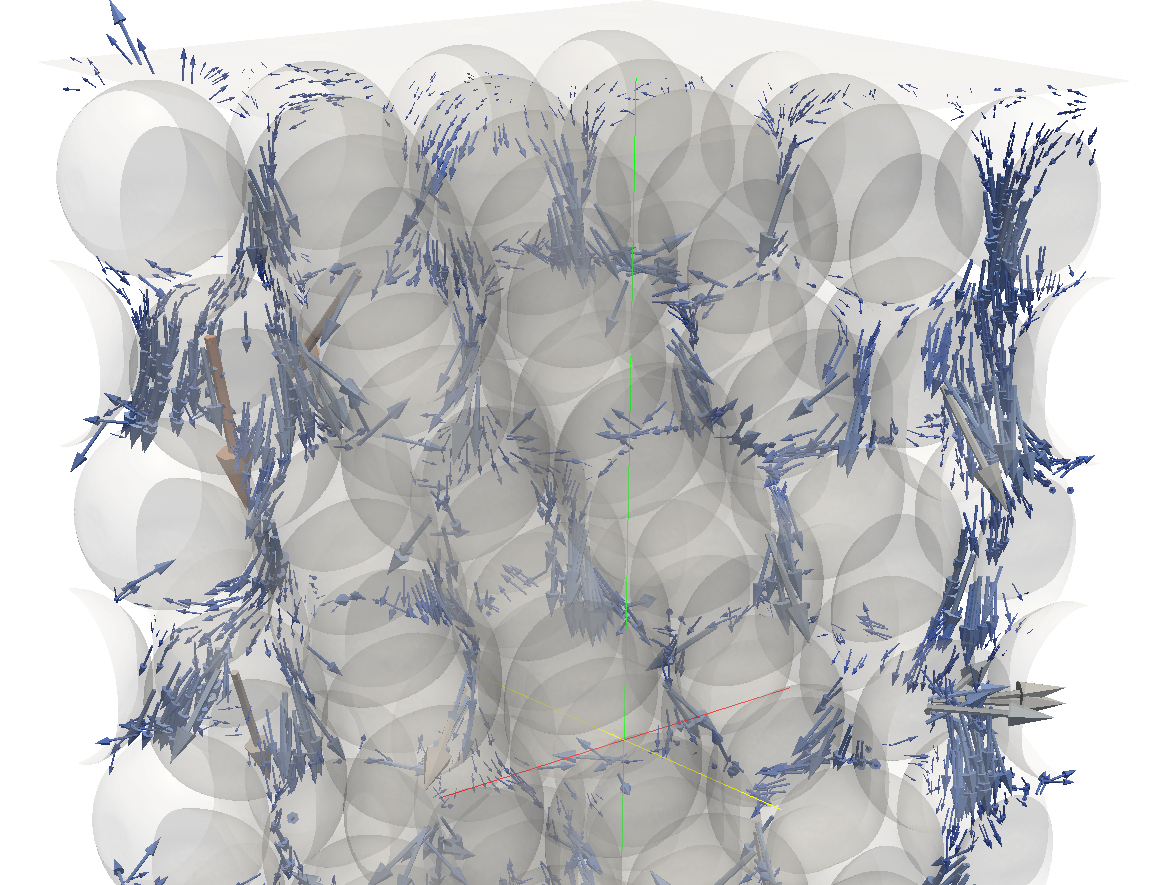}	
		\caption{Simulation of foaming processes in the food industry.}
		\label{fig:foam}
	\end{subfigure}~\hfill
	\begin{subfigure}[t]{0.48\textwidth}	
		\centering
		\includegraphics[width=\textwidth]{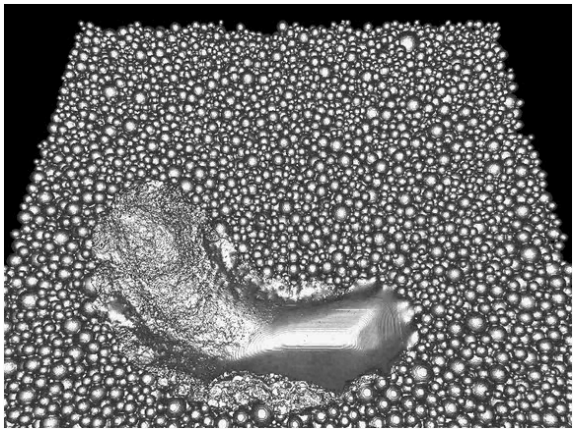}	
		\caption{Simulation of an electron beam that melts a powder bed in additive manufacturing.}
		\label{fig:ebm}
	\end{subfigure}
	\caption{Example of simulations performed using \walberla's free surface lattice Boltzmann extension.}
\end{figure}

The free surface lattice Boltzmann method~(FSLBM) presented in \cite{Koerner2005} is designed for simulating two-phase flow scenarios which are assumed to be only driven by the flow dynamics of one fluid phase.
For instance, in foaming processes with a liquid and gas phase, the phases' viscosities and densities differ by orders of magnitude.
Thus, the influence of the flow in gas phase is considered negligible such that the problem can be treated as a one-phase flow scenario.
The FSLBM extension of \walberla{} has been optimized for parallel computing~\cite{Donath2009,Anderl2014_bubble} and was validated thoroughly~\cite{Donath2011}.
The simulations performed with the FSLBM have already provided new insights into a wide variety of physical applications.
For instance, \walberla's FSLBM extension was used to investigate complicated foaming processes that occur in the food industry~\cite{Anderl2014_foam,Anderl2014_shear} (cf.\ \Cref{fig:foam}).
Furthermore, together with an advection-diffusion LBM implementation, it has been used to simulate an electron beam that melts a powder bed in additive manufacturing~\cite{Ammer2014,Markl2015}, as shown in \Cref{fig:ebm}.

\subsection{Phase-field methods}

\begin{figure}[ht]
\centering
\includegraphics[width=\textwidth]{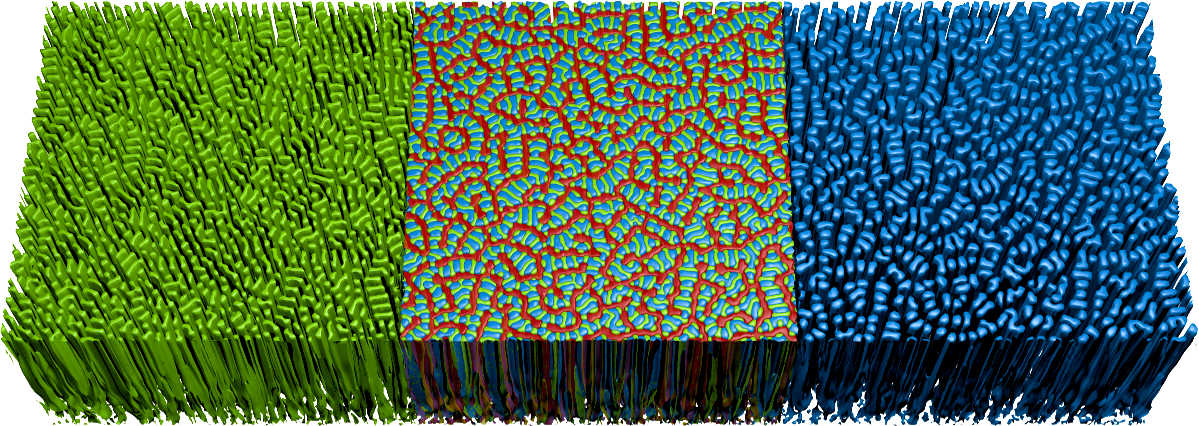}
\caption{Three dimensional simulation of directional solidification of ternary eutectic system Ag-Al-Cu on Hornet supercomputer using $2420 \times 2420 \times 1474$ cells.}
\end{figure}

\walberla{} is used to simulate microstructure formation in metal alloys based on an advanced, thermodynamically consistent phase-field model of Allen-Cahn type~\cite{hoetzer15phasefield}.
These microstructures influence the macroscopic properties of the material significantly. 
Because very large domain sizes are required to observe the formation of microstructural patterns, a HPC implementation is crucial to get physically meaningful results for the considered scenarios. 

In Ref.~\citenum{bauer15phasefield}, we manually optimize and parallelize a basic implementation of the model using \walberla{}'s infrastructure. Buffering techniques, explicit SIMD vectorization, and communication hiding result in a single-node speedup of a factor of 80 compared to a basic C implementation of this model. 
With code generation techniques and an abstract problem formulation in terms of a continuous free energy functional, we can automate these optimizations such that they can be applied to a wide range of different phase-field models~\cite{bauer19phasefield}.
These phase-field application make full use of \walberla{}'s parallelization infrastructure for CPU and GPU clusters.

\section{Conclusion}
\label{sec:conclusion}

In this work, we have presented the \walberla{} multiphysics simulation framework. 
Initially, the framework has been designed as a computational fluid dynamics (CFD) framework based on the lattice Boltzmann method. Over time, it has become a modular HPC toolkit that forms the basis of simulations for a variety of different applications, ranging from carbon nanotubes to additive manufacturing.
A block-structured domain partitioning approach allows for the flexible treatment of complex geometries and can fully exploit modern hardware by utilizing various levels of parallelism.
Its explicit focus on scalability and performance enables efficient usage of HPC systems which grants detailed insight into physical processes of various fluid, particle, and coupled fluid-particle applications.

To ensure a sustainable software development process, \walberla{} uses a continuous integration workflow with extensive unit, integration, and performance tests. This guarantees feature and performance stability for its users while encouraging developers to make changes.

To achieve high node-level performance, compute-intensive parts of the framework can be automatically generated: A high-level description is processed by a set of optimizing transformations and transformed into highly efficient CPU or GPU code. This approach makes \walberla{} ready for future HPC architectures, and permits easy integration of new physical models.

\section*{Acknowledgements}
\label{}
The authors thank
Daniela Anderl, 
Dominik Bartuschat, 
Silke Bergler, 
Simon Bogner, 
Paulo Carvalho, 
Regina Degenhardt, 
Frank Deserno, 
Sagar Dolas, 
Stefan Donath, 
Ehsan Fattahi, 
Christian Feichtinger, 
Jan Götz, 
Johannes Habich, 
Lorenz Hufnagel, 
Klaus Iglberger, 
Sunil Kontham, 
Tobias Leemann, 
Matthias Markl, 
Arash Partow, 
Kristina Pickl, 
Thomas Pohl, 
Tobias Preclik, 
João Victor Tozatti Risso, 
Daniel Ritter, 
Tobias Scharpff, 
Tobias Schruff, 
Dominik Schuster, 
David Staubach, 
Cameron Stewart, 
Nils Thürey,
Rudolf Weeber, 
Lukas Werner and 
Felix Winterhalter
for their contributions that led to the current version of \walberla.

As \walberla{} is designed for massively parallel high performance computing, access to supercomputing facilities is of essential importance.
For providing compute time on such systems, the authors are grateful to the 
Global Scientific Information and Computing Center in Tokyo, 
the Höchstleistungsrechenzentrum Stuttgart, 
the Jülich Supercomputing Centre, 
the Leibniz-Rechenzentrum in Garching, 
the Regionales Rechenzentrum Erlangen and 
the Swiss National Supercomputing Centre in Lugano.

The authors are also grateful for their funding received through various projects, such as project KONWIHR-IV from the Bayerisches Staatsministerium für Wissenschaft und Kunst; projects HPC2SE (01ICH16003D) and SKAMPY (01IH15003A) from the Bundesministerium für Bildung und Forschung (BMBF); projects RU 422/16-2, RU 422/27-1, and SPP 1726 (HO 1108/24-2) from the Deutsche Forschungsgemeinschaft (DFG); and project AiF 17125 from the Arbeitsgemeinschaft industrieller Forschungsvereinigungen (AiF) funded by the Bundesministerium für Wirtschaft und Energie (BMWi).


\section*{References}
\bibliographystyle{elsarticle-num}
\bibliography{Library}	

\begin{thebibliography}{100}
\expandafter\ifx\csname url\endcsname\relax
  \def\url#1{\texttt{#1}}\fi
\expandafter\ifx\csname urlprefix\endcsname\relax\def\urlprefix{URL }\fi
\expandafter\ifx\csname href\endcsname\relax
  \def\href#1#2{#2} \def\path#1{#1}\fi

\bibitem{keyes2013multiphysics}
D.~E. Keyes, L.~C. McInnes, C.~Woodward, W.~Gropp, E.~Myra, M.~Pernice,
  J.~Bell, J.~Brown, A.~Clo, J.~Connors, et~al., Multiphysics simulations:
  Challenges and opportunities, The International Journal of High Performance
  Computing Applications 27~(1) (2013) 4--83.
\newblock \href {http://dx.doi.org/10.1177/1094342012468181}
  {\path{doi:10.1177/1094342012468181}}.

\bibitem{RuedeCSE2016}
U.~R{\"u}de, K.~Willcox, L.~McInnes, H.~Sterck, Research and education in
  computational science and engineering, SIAM Review 60~(3) (2018) 707--754.
\newblock \href {http://dx.doi.org/10.1137/16M1096840}
  {\path{doi:10.1137/16M1096840}}.

\bibitem{godenschwager_framework_2013}
C.~Godenschwager, F.~Schornbaum, M.~Bauer, H.~Köstler, U.~Rüde, A framework
  for hybrid parallel flow simulations with a trillion cells in complex
  geometries, in: Proceedings of the {International} {Conference} on {High}
  {Performance} {Computing}, {Networking}, {Storage} and {Analysis}, ACM Press,
  2013, pp. 1--12.
\newblock \href {http://dx.doi.org/10.1145/2503210.2503273}
  {\path{doi:10.1145/2503210.2503273}}.

\bibitem{feichtinger_2011}
C.~Feichtinger, S.~Donath, H.~Köstler, J.~Götz, U.~Rüde, Walberla: {HPC}
  software design for computational engineering simulations, Journal of
  Computational Science 2~(2) (2011) 105 -- 112.
\newblock \href {http://dx.doi.org/10.1016/j.jocs.2011.01.004}
  {\path{doi:10.1016/j.jocs.2011.01.004}}.

\bibitem{MPI}
{Message Passing Interface}, \url{https://www.mpi-forum.org/}, accessed on
  2019-09-30.

\bibitem{Heuveline2007}
V.~Heuveline, J.~Latt, The {O}pen{LB} project: An open source and object
  oriented implementation of lattice boltzmann methods, International Journal
  of Modern Physics C 18~(04) (2007) 627--634.
\newblock \href {http://dx.doi.org/10.1142/S0129183107010875}
  {\path{doi:10.1142/S0129183107010875}}.

\bibitem{OpenLB}
OpenLB, \url{https://www.openlb.net/}, accessed on 2019-09-30.

\bibitem{Lagrava2012}
D.~Lagrava, O.~Malaspinas, J.~Latt, B.~Chopard, Advances in multi-domain
  lattice boltzmann grid refinement, Journal of Computational Physics 231~(14)
  (2012) 4808 -- 4822.
\newblock \href {http://dx.doi.org/10.1016/j.jcp.2012.03.015}
  {\path{doi:10.1016/j.jcp.2012.03.015}}.

\bibitem{Palabos}
Palabos, \url{http://www.palabos.org/}, accessed on 2019-09-30.

\bibitem{Mierke2018}
D.~Mierke, C.~Janßen, T.~Rung, An efficient algorithm for the calculation of
  sub-grid distances for higher-order lbm boundary conditions in a gpu
  simulation environment, Computers \& Mathematics with Applications\href
  {http://dx.doi.org/10.1016/j.camwa.2018.04.022}
  {\path{doi:10.1016/j.camwa.2018.04.022}}.

\bibitem{elbe}
elbe, \url{https://www.tuhh.de/elbe/home.html}, accessed on 2019-09-30.

\bibitem{Groen2011}
D.~Groen, O.~Henrich, F.~Janoschek, P.~Coveney, J.~Harting, Lattice-boltzmann
  methods in fluid dynamics: Turbulence and complex colloidal fluids, in:
  J{\"u}lich Blue Gene/P Extreme Scaling Workshop, 2011, p.~17.

\bibitem{schmieschek2017}
S.~Schmieschek, L.~Shamardin, S.~Frijters, T.~Kr{\"u}ger, U.~D. Schiller,
  J.~Harting, P.~V. Coveney, {L}{B}3{D}: A parallel implementation of the
  lattice-boltzmann method for simulation of interacting amphiphilic fluids,
  Computer Physics Communications 217 (2017) 149--161.
\newblock \href {http://dx.doi.org/10.1016/j.cpc.2017.03.013}
  {\path{doi:10.1016/j.cpc.2017.03.013}}.

\bibitem{LB3D}
LB3D, \url{http://ccs.chem.ucl.ac.uk/lb3d}, accessed on 2019-09-30.

\bibitem{Groen2013}
D.~Groen, J.~Hetherington, H.~B. Carver, R.~W. Nash, M.~O. Bernabeu, P.~V.
  Coveney, Analysing and modelling the performance of the hemelb
  lattice-boltzmann simulation environment, Journal of Computational Science
  4~(5) (2013) 412 -- 422.
\newblock \href {http://dx.doi.org/10.1016/j.jocs.2013.03.002}
  {\path{doi:10.1016/j.jocs.2013.03.002}}.

\bibitem{Sailfish}
Sailfish, \url{https://github.com/sailfish-team/sailfish}, accessed on
  2019-09-30.

\bibitem{Liu2019}
Z.~{Liu}, X.~{Chu}, X.~{Lv}, H.~{Meng}, S.~{Shi}, W.~{Han}, J.~{Xu}, H.~{Fu},
  G.~{Yang}, Sunwaylb: Enabling extreme-scale lattice boltzmann method based
  computing fluid dynamics simulations on sunway taihulight, in: 2019 IEEE
  International Parallel and Distributed Processing Symposium (IPDPS), 2019,
  pp. 557--566.
\newblock \href {http://dx.doi.org/10.1109/IPDPS.2019.00065}
  {\path{doi:10.1109/IPDPS.2019.00065}}.

\bibitem{wittmann2018}
M.~Wittmann, V.~Haag, T.~Zeiser, H.~K\"ostler, G.~Wellein, Lattice {B}oltzmann
  benchmark kernels as a testbed for performance analysis, Computers \& Fluids
  172 (2018) 582 -- 592.
\newblock \href {http://dx.doi.org/10.1016/j.compfluid.2018.03.030}
  {\path{doi:10.1016/j.compfluid.2018.03.030}}.

\bibitem{LIGGGHTS}
LIGGGHTS, \url{https://www.cfdem.com/}, accessed on 2019-09-30.

\bibitem{GranOO}
GranOO, \url{https://www.yakuru.fr/granoo/index.html}, accessed on 2019-09-30.

\bibitem{YADE}
YADE, \url{https://yade-dev.gitlab.io/trunk/}, accessed on 2019-09-30.

\bibitem{ProjectChrono}
PROJECTCHRONO, \url{https://projectchrono.org/}, accessed on 2019-09-30.

\bibitem{MercuryDPM}
MercuryDPM, \url{http://www.mercurydpm.org/}, accessed on 2019-09-30.

\bibitem{Deiterding2016}
R.~Deiterding, S.~L. Wood, Predictive wind turbine simulation with an adaptive
  lattice boltzmann method for moving boundaries, Journal of Physics:
  Conference Series 753~(8) (2016) 082005.
\newblock \href {http://dx.doi.org/10.1088/1742-6596/753/8/082005}
  {\path{doi:10.1088/1742-6596/753/8/082005}}.

\bibitem{AMROC}
AMROC, \url{https://amroc.sourceforge.net/}, accessed on 2019-09-30.

\bibitem{Burstedde2011}
C.~Burstedde, L.~Wilcox, O.~Ghattas, p4est: Scalable algorithms for parallel
  adaptive mesh refinement on forests of octrees, SIAM Journal on Scientific
  Computing 33~(3) (2011) 1103--1133.
\newblock \href {http://dx.doi.org/10.1137/100791634}
  {\path{doi:10.1137/100791634}}.

\bibitem{Mehl2011}
M.~Mehl, T.~Neckel, P.~Neumann, Navier–stokes and lattice–boltzmann on
  octree-like grids in the peano framework, International Journal for Numerical
  Methods in Fluids 65~(1--3) (2011) 67--86.
\newblock \href {http://dx.doi.org/10.1002/fld.2469}
  {\path{doi:10.1002/fld.2469}}.

\bibitem{WalberlaGit}
{Walberla Gitlab}, \url{https://i10git.cs.fau.de/walberla/walberla/}, accessed
  on 2019-09-30.

\bibitem{walberlaWebsite}
waLBerla, \url{https://www.walberla.net}, accessed on 2019-09-30.

\bibitem{Schornbaum2016}
F.~Schornbaum, U.~R{\"{u}}de, Massively parallel algorithms for the lattice
  boltzmann method on {NonUniform} grids, {SIAM} Journal on Scientific
  Computing 38~(2) (2016) 96--126.
\newblock \href {http://dx.doi.org/10.1137/15M1035240}
  {\path{doi:10.1137/15M1035240}}.

\bibitem{Schornbaum2018}
F.~Schornbaum, U.~R{\"u}de, Extreme-scale block-structured adaptive mesh
  refinement, {SIAM} Journal on Scientific Computing 40~(3) (2018) 358--387.
\newblock \href {http://dx.doi.org/10.1137/17m1128411}
  {\path{doi:10.1137/17m1128411}}.

\bibitem{Dubey2014}
A.~Dubey, A.~Almgren, J.~Bell, M.~Berzins, S.~Brandt, G.~Bryan, P.~Colella,
  D.~Graves, M.~Lijewski, F.~Löffler, B.~O’Shea, E.~Schnetter, B.~V.
  Straalen, K.~Weide, A survey of high level frameworks in block-structured
  adaptive mesh refinement packages, Journal of Parallel and Distributed
  Computing 74~(12) (2014) 3217 -- 3227.
\newblock \href {http://dx.doi.org/10.1016/j.jpdc.2014.07.001}
  {\path{doi:10.1016/j.jpdc.2014.07.001}}.

\bibitem{Schloegel2002}
K.~Schloegel, G.~Karypis, V.~Kumar, Parallel static and dynamic
  multi-constraint graph partitioning, Concurrency and Computation: Practice
  and Experience 14~(3) (2002) 219--240.
\newblock \href {http://dx.doi.org/10.1002/cpe.605}
  {\path{doi:10.1002/cpe.605}}.

\bibitem{Parmetis}
ParMETIS, \url{http://glaros.dtc.umn.edu/gkhome/views/metis/}, accessed on
  2019-09-30.

\bibitem{OpenMP}
OpenMP, \url{https://www.openmp.org/}, accessed on 2019-09-30.

\bibitem{TOP500}
TOP500, \url{https://www.top500.org/}, accessed on 2019-09-30.

\bibitem{Snir2014}
M.~Snir, R.~W. Wisniewski, J.~A. Abraham, S.~V. Adve, S.~Bagchi, P.~Balaji,
  J.~Belak, P.~Bose, F.~Cappello, B.~Carlson, A.~A. Chien, P.~Coteus, N.~A.
  DeBardeleben, P.~C. Diniz, C.~Engelmann, M.~Erez, S.~Fazzari, A.~Geist,
  R.~Gupta, F.~Johnson, S.~Krishnamoorthy, S.~Leyffer, D.~Liberty, S.~Mitra,
  T.~Munson, R.~Schreiber, J.~Stearley, E.~V. Hensbergen, Addressing failures
  in exascale computing, The International Journal of High Performance
  Computing Applications 28~(2) (2014) 129--173.
\newblock \href {http://dx.doi.org/10.1177/1094342014522573}
  {\path{doi:10.1177/1094342014522573}}.

\bibitem{Dongarra2013}
J.~Dongarra, Emerging heterogeneous technologies for high performance
  computing,
  \url{http://www.netlib.org/utk/people/JackDongarra/SLIDES/hcw-0513.pdf},
  accessed on 2019-09-30.

\bibitem{Huang1984}
{Kuang-Hua Huang}, J.~A. {Abraham}, Algorithm-based fault tolerance for matrix
  operations, IEEE Transactions on Computers C-33~(6) (1984) 518--528.
\newblock \href {http://dx.doi.org/10.1109/TC.1984.1676475}
  {\path{doi:10.1109/TC.1984.1676475}}.

\bibitem{Randell1975}
B.~{Randell}, System structure for software fault tolerance, IEEE Transactions
  on Software Engineering SE-1~(2) (1975) 220--232.
\newblock \href {http://dx.doi.org/10.1109/TSE.1975.6312842}
  {\path{doi:10.1109/TSE.1975.6312842}}.

\bibitem{Huber2016Resilience}
M.~Huber, B.~Gmeiner, U.~R\"ude, B.~Wohlmuth, Resilience for massively parallel
  multigrid solvers, SIAM J.~Sci.~Comp. 38~(5) (2016) S217--S239.
\newblock \href {http://dx.doi.org/10.1137/15M1026122}
  {\path{doi:10.1137/15M1026122}}.

\bibitem{zheng2012scalable}
G.~Zheng, X.~Ni, L.~V. Kal{\'e}, A scalable double in-memory checkpoint and
  restart scheme towards exascale, in: IEEE/IFIP International Conference on
  Dependable Systems and Networks Workshops (DSN 2012), IEEE, 2012, pp. 1--6.
\newblock \href {http://dx.doi.org/10.1109/DSNW.2012.6264677}
  {\path{doi:10.1109/DSNW.2012.6264677}}.

\bibitem{herault2015fault}
T.~Herault, Y.~Robert, Fault-tolerance techniques for high-performance
  computing, Springer, 2015.
\newblock \href {http://dx.doi.org/10.1007/978-3-319-20943-2}
  {\path{doi:10.1007/978-3-319-20943-2}}.

\bibitem{Kohl_2019}
N.~Kohl, J.~Hötzer, F.~Schornbaum, M.~Bauer, C.~Godenschwager, H.~Köstler,
  B.~Nestler, U.~Rüde, A scalable and extensible checkpointing scheme for
  massively parallel simulations, The International Journal of High Performance
  Computing Applications 33~(4) (2019) 571--589.
\newblock \href {http://dx.doi.org/10.1177/1094342018767736}
  {\path{doi:10.1177/1094342018767736}}.

\bibitem{Bland2013ULFM}
W.~Bland, A.~Bouteiller, T.~Herault, G.~Bosilca, J.~Dongarra, Post-failure
  recovery of {MPI} communication capability: Design and rationale,
  International Journal of High Performance Computing Applications 27 (2013)
  244 -- 254.
\newblock \href {http://dx.doi.org/10.1177/1094342013488238}
  {\path{doi:10.1177/1094342013488238}}.

\bibitem{lorensen1987marching}
W.~E. Lorensen, H.~E. Cline, Marching cubes: A high resolution {3D} surface
  construction algorithm, in: ACM siggraph computer graphics, Vol.~21, ACM,
  1987, pp. 163--169.
\newblock \href {http://dx.doi.org/10.1145/37402.37422}
  {\path{doi:10.1145/37402.37422}}.

\bibitem{bauer2015massively}
M.~Bauer, J.~H{\"o}tzer, M.~Jainta, P.~Steinmetz, M.~Berghoff, F.~Schornbaum,
  C.~Godenschwager, H.~K{\"o}stler, B.~Nestler, U.~R{\"u}de, Massively parallel
  phase-field simulations for ternary eutectic directional solidification, in:
  SC'15: Proceedings of the International Conference for High Performance
  Computing, Networking, Storage and Analysis, IEEE, 2015, pp. 1--12.
\newblock \href {http://dx.doi.org/10.1145/2807591.2807662}
  {\path{doi:10.1145/2807591.2807662}}.

\bibitem{garland1997surface}
M.~Garland, P.~S. Heckbert, Surface simplification using quadric error metrics,
  in: Proceedings of the 24th annual conference on Computer graphics and
  interactive techniques, ACM Press/Addison-Wesley Publishing Co., 1997, pp.
  209--216.
\newblock \href {http://dx.doi.org/10.1145/258734.258849}
  {\path{doi:10.1145/258734.258849}}.

\bibitem{OpenMesh}
OpenMesh, \url{https://www.openmesh.org/}, accessed on 2019-09-30.

\bibitem{Jones1995}
M.~W. Jones, {3D} distance from a point to a triangle, Tech. rep., Department
  of Computer Science, University of Wales (1995).

\bibitem{Berentzen2005}
J.~B{\ae}rentzen, H.~Aan{\ae}s, Signed distance computation using the angle
  weighted pseudonormal, Visualization and Computer Graphics, IEEE Transactions
  on 11~(3) (2005) 243--253.
\newblock \href {http://dx.doi.org/10.1109/TVCG.2005.49}
  {\path{doi:10.1109/TVCG.2005.49}}.

\bibitem{Payne1992}
B.~Payne, A.~Toga, Distance field manipulation of surface models, Computer
  Graphics and Applications, IEEE 12~(1) (1992) 65--71.
\newblock \href {http://dx.doi.org/10.1109/38.135885}
  {\path{doi:10.1109/38.135885}}.

\bibitem{Krueger2017}
T.~Kr{\"u}ger, H.~Kusumaatmaja, A.~Kuzmin, O.~Shardt, G.~Silva, E.~M. Viggen,
  The lattice Boltzmann method, Springer, 2017.
\newblock \href {http://dx.doi.org/10.1007/978-3-319-44649-3}
  {\path{doi:10.1007/978-3-319-44649-3}}.

\bibitem{dHumieres2002}
D.~d'Humieres, Multiple--relaxation--time lattice boltzmann models in three
  dimensions, Philosophical Transactions of the Royal Society of London. Series
  A: Mathematical, Physical and Engineering Sciences 360~(1792) (2002)
  437--451.
\newblock \href {http://dx.doi.org/10.1098/rsta.2001.0955}
  {\path{doi:10.1098/rsta.2001.0955}}.

\bibitem{Ginzburg2008}
I.~Ginzburg, F.~Verhaeghe, D.~d'Humieres, Two-relaxation-time lattice
  {Boltzmann} scheme: {About} parametrization, velocity, pressure and mixed
  boundary conditions, Communications in Computational Physics 3~(2) (2008)
  427--478.

\bibitem{Geier2015}
M.~Geier, M.~Schönherr, A.~Pasquali, M.~Krafczyk, The cumulant lattice
  boltzmann equation in three dimensions: Theory and validation, Computers \&
  Mathematics with Applications 70~(4) (2015) 507 -- 547.
\newblock \href {http://dx.doi.org/10.1016/j.camwa.2015.05.001}
  {\path{doi:10.1016/j.camwa.2015.05.001}}.

\bibitem{Yu2005}
H.~Yu, S.~S. Girimaji, L.-S. Luo, Dns and les of decaying isotropic turbulence
  with and without frame rotation using lattice boltzmann method, Journal of
  Computational Physics 209~(2) (2005) 599 -- 616.
\newblock \href {http://dx.doi.org/10.1016/j.jcp.2005.03.022}
  {\path{doi:10.1016/j.jcp.2005.03.022}}.

\bibitem{KBC2015}
F.~B\"osch, S.~S. Chikatamarla, I.~V. Karlin, Entropic multirelaxation lattice
  boltzmann models for turbulent flows, Phys. Rev. E 92 (2015) 043309.
\newblock \href {http://dx.doi.org/10.1103/PhysRevE.92.043309}
  {\path{doi:10.1103/PhysRevE.92.043309}}.

\bibitem{Junk2008}
M.~Junk, Z.~Yang, Outflow boundary conditions for the lattice boltzmann method,
  Progress in Computational Fluid Dynamics, an International Journal 8~(1-4)
  (2008) 38--48.
\newblock \href {http://dx.doi.org/10.1504/PCFD.2008.018077}
  {\path{doi:10.1504/PCFD.2008.018077}}.

\bibitem{Guo2002}
Z.~Guo, C.~Zheng, B.~Shi, Discrete lattice effects on the forcing term in the
  lattice boltzmann method, Phys. Rev. E 65 (2002) 046308.
\newblock \href {http://dx.doi.org/10.1103/PhysRevE.65.046308}
  {\path{doi:10.1103/PhysRevE.65.046308}}.

\bibitem{Rohde2006}
M.~Rohde, D.~Kandhai, J.~J. Derksen, H.~E.~A. van~den Akker, A generic, mass
  conservative local grid refinement technique for lattice-boltzmann schemes,
  International Journal for Numerical Methods in Fluids 51~(4) (2006) 439--468.
\newblock \href {http://dx.doi.org/10.1002/fld.1140}
  {\path{doi:10.1002/fld.1140}}.

\bibitem{zeiser2008introducing}
T.~Zeiser, G.~Wellein, A.~Nitsure, K.~Iglberger, U.~Rüde, G.~Hager,
  Introducing a parallel cache oblivious blocking approach for the lattice
  boltzmann method, Progress in Computational Fluid Dynamics, an International
  Journal 8~(1-4) (2008) 179--188.
\newblock \href {http://dx.doi.org/10.1504/PCFD.2008.018088}
  {\path{doi:10.1504/PCFD.2008.018088}}.

\bibitem{donath2008performance}
S.~Donath, K.~Iglberger, G.~Wellein, T.~Zeiser, A.~Nitsure, U.~Rüde,
  Performance comparison of different parallel lattice boltzmann
  implementations on multi-core multi-socket systems, International Journal of
  Computational Science and Engineering 4~(1) (2008) 3--11.
\newblock \href {http://dx.doi.org/10.1504/IJCSE.2008.021107}
  {\path{doi:10.1504/IJCSE.2008.021107}}.

\bibitem{wellein2006single}
G.~Wellein, T.~Zeiser, G.~Hager, S.~Donath, On the single processor performance
  of simple lattice boltzmann kernels, Computers \& Fluids 35~(8-9) (2006)
  910--919.
\newblock \href {http://dx.doi.org/10.1016/j.compfluid.2005.02.008}
  {\path{doi:10.1016/j.compfluid.2005.02.008}}.

\bibitem{HighQClub}
{High-Q Club J\"ulich},
  \url{http://www.fz-juelich.de/ias/jsc/EN/Expertise/High-Q-Club/_node.html},
  accessed on 2019-09-30.

\bibitem{Gil2017}
A.~Gil, J.~Galache, C.~Godenschwager, U.~Rüde, Optimum configuration for
  accurate simulations of chaotic porous media with lattice boltzmann methods
  considering boundary conditions, lattice spacing and domain size, Computers
  \& Mathematics with Applications 73~(12) (2017) 2515 -- 2528.
\newblock \href {http://dx.doi.org/10.1016/j.camwa.2017.03.017}
  {\path{doi:10.1016/j.camwa.2017.03.017}}.

\bibitem{Fattahi_etal_2016}
E.~Fattahi, C.~Waluga, B.~Wohlmuth, U.~R\"ude, Large scale lattice {B}oltzmann
  simulation for the coupling of free and porous media flow, in: Proceedings of
  the {I}nternational {C}onference on {H}igh {P}erformance {C}omputing in
  {S}cience and {E}ngineering, Springer, 2016, pp. 1--18.
\newblock \href {http://dx.doi.org/10.1007/978-3-319-40361-8_1}
  {\path{doi:10.1007/978-3-319-40361-8_1}}.

\bibitem{Fattahi_etal_2016b}
E.~Fattahi, C.~Waluga, B.~Wohlmuth, U.~R\"ude, M.~Manhart, R.~Helmig, Lattice
  {B}oltzmann methods in porous media simulations: {F}rom laminar to turbulent
  flow, Computers \& Fluids 140 (2016) 247--259.
\newblock \href {http://dx.doi.org/10.1016/j.compfluid.2016.10.007}
  {\path{doi:10.1016/j.compfluid.2016.10.007}}.

\bibitem{Rybak_2019}
I.~Rybak, C.~Schwarzmeier, E.~Eggenweiler, U.~R\"ude, Validation and
  calibration of coupled porous-medium and free-flow problems using pore-scale
  resolved models, submitted manuscript: \url{https://arxiv.org/abs/1906.06884}
  (2019).

\bibitem{Eibl2019}
S.~Eibl, U.~R{\"u}de, A systematic comparison of runtime load balancing
  algorithms for massively parallel rigid particle dynamics, Computer Physics
  Communications 244 (2019) 76--85.
\newblock \href {http://dx.doi.org/10.1016/j.cpc.2019.06.020}
  {\path{doi:10.1016/j.cpc.2019.06.020}}.

\bibitem{Hockney1974}
R.~Hockney, S.~Goel, J.~Eastwood, Quiet high-resolution computer models of a
  plasma, Journal of Computational Physics 14~(2) (1974) 148--158.
\newblock \href {http://dx.doi.org/10.1016/0021-9991(74)90010-2}
  {\path{doi:10.1016/0021-9991(74)90010-2}}.

\bibitem{Allen2017}
M.~P. Allen, D.~J. Tildesley, Computer simulation of liquids, Oxford university
  press, 2017.

\bibitem{Ericson2004}
C.~Ericson, Real-time collision detection, CRC Press, 2004.

\bibitem{Erleben2005}
K.~Erleben, J.~Sporring, K.~Henriksen, K.~Dohlman, Physics-based animation
  (graphics series) (2005).

\bibitem{Gilbert1988}
E.~Gilbert, D.~Johnson, S.~Keerthi, A fast procedure for computing the distance
  between complex objects in three-dimensional space, {IEEE} Journal on
  Robotics and Automation 4~(2) (1988) 193--203.
\newblock \href {http://dx.doi.org/10.1109/56.2083}
  {\path{doi:10.1109/56.2083}}.

\bibitem{Gilbert1990}
E.~G. Gilbert, C.-P. Foo, Computing the distance between general convex objects
  in three-dimensional space, IEEE Transactions on Robotics and Automation
  6~(1) (1990) 53--61.
\newblock \href {http://dx.doi.org/10.1109/70.88117}
  {\path{doi:10.1109/70.88117}}.

\bibitem{Bergen2003}
G.~V.~D. Bergen, Collision Detection in Interactive {3D} Environments, FOCAL
  PR, 2003.

\bibitem{CUNDALL1971}
P.~A. Cundall, A computer model for simulating progressive, large-scale
  movements in blocky rock systems, Proc. Int. Symp. on Rock Fracture (1971)
  11--8.

\bibitem{Cundall1979}
P.~A. Cundall, O.~D.~L. Strack, A discrete numerical model for granular
  assemblies, G{\'{e}}otechnique 29~(1) (1979) 47--65.
\newblock \href {http://dx.doi.org/10.1680/geot.1979.29.1.47}
  {\path{doi:10.1680/geot.1979.29.1.47}}.

\bibitem{Preclik2015}
T.~Preclik, U.~R{\"{u}}de, Ultrascale simulations of non-smooth granular
  dynamics, Computational Particle Mechanics 2~(2) (2015) 173--196.
\newblock \href {http://dx.doi.org/10.1007/s40571-015-0047-6}
  {\path{doi:10.1007/s40571-015-0047-6}}.

\bibitem{Preclik2017}
T.~Preclik, S.~Eibl, U.~R{\"{u}}de, The maximum dissipation principle in
  rigid-body dynamics with inelastic impacts, Computational Mechanics 62~(1)
  (2017) 1--16.
\newblock \href {http://dx.doi.org/10.1007/s00466-017-1486-0}
  {\path{doi:10.1007/s00466-017-1486-0}}.

\bibitem{Rapaport1991}
D.~Rapaport, Multi-million particle molecular dynamics: Ii. design
  considerations for distributed processing, Computer Physics Communications
  62~(2-3) (1991) 217--228.
\newblock \href {http://dx.doi.org/10.1016/0010-4655(91)90096-4}
  {\path{doi:10.1016/0010-4655(91)90096-4}}.

\bibitem{Eibl2018}
S.~Eibl, U.~R{\"{u}}de, A local parallel communication algorithm for
  polydisperse rigid body dynamics, Parallel Computing 80 (2018) 36--48.
\newblock \href {http://dx.doi.org/10.1016/j.parco.2018.10.002}
  {\path{doi:10.1016/j.parco.2018.10.002}}.

\bibitem{Schruff2016}
T.~Schruff, R.~Liang, U.~R{\"u}de, H.~Sch{\"u}ttrumpf, R.~M. Frings, Generation
  of dense granular deposits for porosity analysis: assessment and application
  of large-scale non-smooth granular dynamics, Computational Particle Mechanics
  5~(1) (2016) 1--12.
\newblock \href {http://dx.doi.org/10.1007/s40571-016-0153-0}
  {\path{doi:10.1007/s40571-016-0153-0}}.

\bibitem{Ostanin2018}
I.~A. Ostanin, P.~Zhilyaev, V.~Petrov, T.~Dumitrica, S.~Eibl, U.~Rüde, V.~A.
  Kuzkin, Toward large scale modeling of carbon nanotube systems with the
  mesoscopic distinct element method, Letters on Materials 8~(3) (2018)
  240--245.
\newblock \href {http://dx.doi.org/10.22226/2410-3535-2018-3-240-245}
  {\path{doi:10.22226/2410-3535-2018-3-240-245}}.

\bibitem{Ostanin2019}
I.~Ostanin, T.~Dumitrica, S.~Eibl, U.~Rüde, Size-independent mechanical
  response of ultrathin {CNT} films in mesoscopic distinct element method
  simulations, Journal of Applied Mechanics (2019) 1--17\href
  {http://dx.doi.org/10.1115/1.4044413} {\path{doi:10.1115/1.4044413}}.

\bibitem{Ladd1994}
A.~J.~C. Ladd, Numerical simulations of particulate suspensions via a
  discretized boltzmann equation. part 1. theoretical foundation, Journal of
  Fluid Mechanics 271 (1994) 285–309.
\newblock \href {http://dx.doi.org/10.1017/S0022112094001771}
  {\path{doi:10.1017/S0022112094001771}}.

\bibitem{Aidun1998}
C.~K. Aidun, Y.~Lu, E.-J. Ding, Direct analysis of particulate suspensions with
  inertia using the discrete boltzmann equation, Journal of Fluid Mechanics 373
  (1998) 287–311.
\newblock \href {http://dx.doi.org/10.1017/S0022112098002493}
  {\path{doi:10.1017/S0022112098002493}}.

\bibitem{Noble1998}
D.~R. Noble, J.~R. Torczynski, A {Lattice}-{Boltzmann} {Method} for {Partially}
  {Saturated} {Computational} {Cells}, International Journal of Modern Physics
  C 09~(08) (1998) 1189--1201.
\newblock \href {http://dx.doi.org/10.1142/S0129183198001084}
  {\path{doi:10.1142/S0129183198001084}}.

\bibitem{Rettinger2017}
C.~Rettinger, U.~R{\"u}de, A comparative study of fluid-particle coupling
  methods for fully resolved lattice boltzmann simulations, Computers \& Fluids
  154 (2017) 74 -- 89.
\newblock \href {http://dx.doi.org/10.1016/j.compfluid.2017.05.033}
  {\path{doi:10.1016/j.compfluid.2017.05.033}}.

\bibitem{Zou1997}
Q.~Zou, X.~He, On pressure and velocity boundary conditions for the lattice
  boltzmann bgk model, Physics of Fluids 9~(6) (1997) 1591--1598.
\newblock \href {http://dx.doi.org/10.1063/1.869307}
  {\path{doi:10.1063/1.869307}}.

\bibitem{Peng2016}
C.~Peng, Y.~Teng, B.~Hwang, Z.~Guo, L.-P. Wang, Implementation issues and
  benchmarking of lattice boltzmann method for moving rigid particle
  simulations in a viscous flow, Computers \& Mathematics with Applications
  72~(2) (2016) 349 -- 374.
\newblock \href {http://dx.doi.org/10.1016/j.camwa.2015.08.027}
  {\path{doi:10.1016/j.camwa.2015.08.027}}.

\bibitem{Rettinger2019}
C.~Rettinger, U.~R{\"u}de, Dynamic load balancing techniques for particulate
  flow simulations, Computation 7~(1).
\newblock \href {http://dx.doi.org/10.3390/computation7010009}
  {\path{doi:10.3390/computation7010009}}.

\bibitem{RettingerRiverbed2017}
C.~Rettinger, C.~Godenschwager, S.~Eibl, T.~Preclik, T.~Schruff, R.~Frings,
  U.~R{\"u}de, Fully resolved simulations of dune formation in riverbeds, in:
  J.~M. Kunkel, R.~Yokota, P.~Balaji, D.~Keyes (Eds.), High Performance
  Computing, Springer International Publishing, Cham, 2017, pp. 3--21.
\newblock \href {http://dx.doi.org/10.1007/978-3-319-58667-0_1}
  {\path{doi:10.1007/978-3-319-58667-0_1}}.

\bibitem{Huang_2004}
L.~R. Huang, E.~C. Cox, R.~H. Austin, J.~C. Sturm, Continuous particle
  separation through deterministic lateral displacement, Science 304~(5673)
  (2004) 987--990.
\newblock \href {http://dx.doi.org/10.1126/science.1094567}
  {\path{doi:10.1126/science.1094567}}.

\bibitem{McGrath2014}
J.~McGrath, M.~Jimenez, H.~Bridle, Deterministic lateral displacement for
  particle separation: a review, Lab Chip 14 (2014) 4139--4158.
\newblock \href {http://dx.doi.org/10.1039/C4LC00939H}
  {\path{doi:10.1039/C4LC00939H}}.

\bibitem{Kuron_etal_2019a}
M.~Kuron, P.~St{\"a}rk, C.~Burkard, J.~de~Graaf, C.~Holm, A lattice boltzmann
  model for squirmers, The Journal of Chemical Physics 150~(14) (2019) 144110.
\newblock \href {http://dx.doi.org/10.1063/1.5085765}
  {\path{doi:10.1063/1.5085765}}.

\bibitem{Kuron_etal_2019b}
M.~Kuron, P.~St{\"a}rk, C.~Holm, J.~de~Graaf, Hydrodynamic mobility reversal of
  squirmers near flat and curved surfaces, Soft Matter 15 (2019) 5908--5920.
\newblock \href {http://dx.doi.org/10.1039/C9SM00692C}
  {\path{doi:10.1039/C9SM00692C}}.

\bibitem{Elgeti2015}
J.~Elgeti, R.~G. Winkler, G.~Gompper, Physics of microswimmers—single
  particle motion and collective behavior: a review, Reports on Progress in
  Physics 78~(5) (2015) 056601.
\newblock \href {http://dx.doi.org/10.1088/0034-4885/78/5/056601}
  {\path{doi:10.1088/0034-4885/78/5/056601}}.

\bibitem{Blake1971}
J.~R. Blake, A spherical envelope approach to ciliary propulsion, Journal of
  Fluid Mechanics 46~(1) (1971) 199--208.
\newblock \href {http://dx.doi.org/10.1017/S002211207100048X}
  {\path{doi:10.1017/S002211207100048X}}.

\bibitem{Lighthill1952}
M.~Lighthill, On the squirming motion of nearly spherical deformable bodies
  through liquids at very small reynolds numbers, Communications on Pure and
  Applied Mathematics 5~(2) (1952) 109--118.
\newblock \href {http://dx.doi.org/10.1002/cpa.3160050201}
  {\path{doi:10.1002/cpa.3160050201}}.

\bibitem{Rettinger2018}
C.~Rettinger, U.~R{\"u}de, A coupled lattice boltzmann method and discrete
  element method for discrete particle simulations of particulate flows,
  Computers \& Fluids 172 (2018) 706 -- 719.
\newblock \href {http://dx.doi.org/10.1016/j.compfluid.2018.01.023}
  {\path{doi:10.1016/j.compfluid.2018.01.023}}.

\bibitem{Schruff2014}
T.~Schruff, F.~Schornbaum, C.~Godenschwager, U.~R{\"u}de, R.~M. Frings,
  H.~Schüttrumpf, Numerical simulation of pore fluid flow and fine sediment
  infiltration into the riverbed, in: 11th International Conference on
  Hydroinformatics, CUNY Academic Works, 2014.

\bibitem{Pippig2013}
M.~Pippig, {P}{F}{F}{T}: An extension of fftw to massively parallel
  architectures, SIAM Journal on Scientific Computing 35~(3) (2013) C213--C236.
\newblock \href {http://dx.doi.org/10.1137/120885887}
  {\path{doi:10.1137/120885887}}.

\bibitem{Bartuschat2015}
D.~Bartuschat, U.~R{\"u}de, Parallel multiphysics simulations of charged
  particles in microfluidic flows, Journal of Computational Science 8 (2015)
  1--19.
\newblock \href {http://dx.doi.org/10.1016/j.jocs.2015.02.006}
  {\path{doi:10.1016/j.jocs.2015.02.006}}.

\bibitem{Capuani2004}
F.~Capuani, I.~Pagonabarraga, D.~Frenkel, Discrete solution of the
  electrokinetic equations, The Journal of Chemical Physics 121~(2) (2004)
  973--986.
\newblock \href {http://dx.doi.org/10.1063/1.1760739}
  {\path{doi:10.1063/1.1760739}}.

\bibitem{Rempfer2016}
G.~Rempfer, G.~B. Davies, C.~Holm, J.~de~Graaf, Reducing spurious flow in
  simulations of electrokinetic phenomena, The Journal of Chemical Physics
  145~(4) (2016) 044901.
\newblock \href {http://dx.doi.org/10.1063/1.4958950}
  {\path{doi:10.1063/1.4958950}}.

\bibitem{Kuron_etal_2016}
M.~Kuron, G.~Rempfer, F.~Schornbaum, M.~Bauer, C.~Godenschwager, C.~Holm,
  J.~de~Graaf, Moving charged particles in lattice boltzmann-based
  electrokinetics, The Journal of Chemical Physics 145~(21) (2016) 214102.
\newblock \href {http://dx.doi.org/10.1063/1.4968596}
  {\path{doi:10.1063/1.4968596}}.

\bibitem{Pystencils}
pystencils, \url{https://i10git.cs.fau.de/pycodegen/pystencils}, accessed on
  2019-09-30.

\bibitem{SymPy}
A.~Meurer, C.~P. Smith, M.~Paprocki, O.~\v{C}ert\'{i}k, S.~B. Kirpichev,
  M.~Rocklin, A.~Kumar, S.~Ivanov, J.~K. Moore, S.~Singh, T.~Rathnayake,
  S.~Vig, B.~E. Granger, R.~P. Muller, F.~Bonazzi, H.~Gupta, S.~Vats,
  F.~Johansson, F.~Pedregosa, M.~J. Curry, A.~R. Terrel, v.~Rou\v{c}ka,
  A.~Saboo, I.~Fernando, S.~Kulal, R.~Cimrman, A.~Scopatz, Sympy: symbolic
  computing in python, PeerJ Computer Science 3 (2017) e103.
\newblock \href {http://dx.doi.org/10.7717/peerj-cs.103}
  {\path{doi:10.7717/peerj-cs.103}}.

\bibitem{PystencilsWalberla}
pystencils~waLBerla interface,
  \url{https://i10git.cs.fau.de/pycodegen/pystencils\_walberla}, accessed on
  2019-09-30.

\bibitem{bauer19phasefield}
M.~Bauer, J.~H{\"o}tzer, D.~Ernst, J.~Hammer, M.~Seitz, H.~Hierl, J.~H\"onig,
  H.~K{\"o}stler, G.~Wellein, B.~Nestler, U.~R{\"u}de, Code generation for
  massively parallel phase-field simulations, in: Proceedings of the
  International Conference for High Performance Computing, Networking, Storage
  and Analysis (accepted, in press), ACM, 2019.

\bibitem{Jinja}
{Jinja Template Language}, \url{https://jinja.palletsprojects.com/}, accessed
  on 2019-09-30.

\bibitem{Eibl2019a}
S.~Eibl, U.~R{\"{u}}de, \href{http://arxiv.org/abs/1906.10963}{A modular and
  extensible software architecture for particle dynamics}, Proceedings of the
  8\textsuperscript{th} International Conference on Discrete Element Methods
  (DEM8).
\newline\urlprefix\url{http://arxiv.org/abs/1906.10963}

\bibitem{Git}
Git, \url{https://git-scm.com/}, accessed on 2019-09-30.

\bibitem{GitLab}
Gitlab, \url{https://gitlab.com/}, accessed on 2019-09-30.

\bibitem{GitHub}
GitHub, \url{https://github.com/}, accessed on 2019-09-30.

\bibitem{Clang}
Clang, \url{https://clang.llvm.org}, accessed on 2019-09-30.

\bibitem{GCC}
{GNU Compiler Collection}, \url{https://gcc.gnu.org}, accessed on 2019-09-30.

\bibitem{Intel}
{Intel C++ Compiler}, \url{https://software.intel.com}, accessed on 2019-09-30.

\bibitem{MSVC}
{Microsoft Visual C++}, \url{https://docs.microsoft.com/cpp}, accessed on
  2019-09-30.

\bibitem{ClangTidy}
Clang-Tidy, \url{https://clang.llvm.org/extra/clang-tidy}, accessed on
  2019-09-30.

\bibitem{Doxygen}
Doxygen, \url{http://www.doxygen.nl/}, accessed on 2019-09-30.

\bibitem{Grafana}
Grafana, \url{https://grafana.com/}, accessed on 2019-09-30.

\bibitem{Docker}
Docker, \url{https://www.docker.com/}, accessed on 2019-09-30.

\bibitem{paraview}
ParaView, \url{https://www.paraview.org/}, accessed on 2019-09-30.

\bibitem{visit}
VisIt, \url{https://wci.llnl.gov/simulation/computer-codes/visit/}, accessed on
  2019-09-30.

\bibitem{bauer2016python}
M.~Bauer, F.~Schornbaum, C.~Godenschwager, M.~Markl, D.~Anderl, H.~K{\"o}stler,
  U.~R{\"u}de, A python extension for the massively parallel multiphysics
  simulation framework walberla, International Journal of Parallel, Emergent
  and Distributed Systems 31~(6) (2016) 529--542.
\newblock \href {http://dx.doi.org/10.1080/17445760.2015.1118478}
  {\path{doi:10.1080/17445760.2015.1118478}}.

\bibitem{BoostPython}
Boost.Python, \url{https://www.boost.org/doc/libs/1\_66\_0/libs/python/},
  accessed on 2019-09-30.

\bibitem{NumPy}
{NumPy}, \url{https://numpy.org/}, accessed on 2019-09-30.

\bibitem{Kohl2019Hyteg}
N.~Kohl, D.~Th{\"o}nnes, D.~Drzisga, D.~Bartuschat, U.~R{\"u}de, The
  {H}y{T}e{G} finite-element software framework for scalable multigrid solvers,
  International Journal of Parallel, Emergent and Distributed Systems 34~(5)
  (2019) 477--496.
\newblock \href {http://dx.doi.org/10.1080/17445760.2018.1506453}
  {\path{doi:10.1080/17445760.2018.1506453}}.

\bibitem{Koerner2005}
C.~Körner, M.~Thies, T.~Hofmann, N.~Thürey, U.~Rüde, Lattice boltzmann model
  for free surface flow for modeling foaming, Journal of Statistical Physics
  121~(1) (2005) 179--196.
\newblock \href {http://dx.doi.org/10.1007/s10955-005-8879-8}
  {\path{doi:10.1007/s10955-005-8879-8}}.

\bibitem{Donath2009}
S.~Donath, C.~Feichtinger, T.~Pohl, J.~Götz, U.~R{\"u}de, Localized parallel
  algorithm for bubble coalescence in free surface lattice-boltzmann method,
  in: H.~Sips, D.~Epema, H.~Lin (Eds.), Euro-Par 2009 Parallel Processing,
  Lecture Notes in Computer Science, vol 5704, Springer, Berlin, Heidelberg,
  Cham, 2009, pp. 735--746.
\newblock \href {http://dx.doi.org/10.1007/978-3-642-03869-3_69}
  {\path{doi:10.1007/978-3-642-03869-3_69}}.

\bibitem{Anderl2014_bubble}
D.~Anderl, S.~Bogner, C.~Rauh, U.~Rüde, A.~Delgado, Free surface lattice
  {Boltzmann} with enhanced bubble model, Computers \& Mathematics with
  Applications 67~(2) (2014) 331--339.
\newblock \href {http://dx.doi.org/10.1016/j.camwa.2013.06.007}
  {\path{doi:10.1016/j.camwa.2013.06.007}}.

\bibitem{Donath2011}
S.~Donath, K.~Mecke, S.~Rabha, V.~Buwa, U.~Rüde, Verification of surface
  tension in the parallel free surface lattice {B}oltzmann method in
  wa{LB}erla, Computers \& Fluids 45~(1) (2011) 177--186.
\newblock \href {http://dx.doi.org/10.1016/j.compfluid.2010.12.027}
  {\path{doi:10.1016/j.compfluid.2010.12.027}}.

\bibitem{Anderl2014_foam}
D.~Anderl, M.~Bauer, C.~Rauh, U.~Rüde, A.~Delgado, Numerical simulation of
  adsorption and bubble interaction in protein foams using a lattice boltzmann
  method, Food \& Function 5 (2014) 755--763.
\newblock \href {http://dx.doi.org/10.1039/C3FO60374A}
  {\path{doi:10.1039/C3FO60374A}}.

\bibitem{Anderl2014_shear}
D.~Anderl, M.~Bauer, C.~Rauh, U.~R{\"u}de, A.~Delgado, Numerical simulation of
  bubbles in shear flow, PAMM 14~(1) (2014) 667--668.
\newblock \href {http://dx.doi.org/10.1002/pamm.201410317}
  {\path{doi:10.1002/pamm.201410317}}.

\bibitem{Ammer2014}
R.~Ammer, M.~Markl, U.~Ljungblad, C.~K{\"o}rner, U.~R{\"u}de, Simulating fast
  electron beam melting with a parallel thermal free surface lattice
  {B}oltzmann method, Computers and Mathematics with Applications 67 (2014)
  318--330.
\newblock \href {http://dx.doi.org/10.1016/j.camwa.2013.10.001}
  {\path{doi:10.1016/j.camwa.2013.10.001}}.

\bibitem{Markl2015}
M.~Markl, R.~Ammer, U.~R{\"u}de, C.~K{\"o}rner, Numerical investigations on
  hatching process strategies for powder-bed-based additive manufacturing using
  an electron beam, The International Journal of Advanced Manufacturing
  Technology 78~(1-4) (2015) 239--247.
\newblock \href {http://dx.doi.org/10.1007/s00170-014-6594-9}
  {\path{doi:10.1007/s00170-014-6594-9}}.

\bibitem{hoetzer15phasefield}
J.~Hötzer, M.~Jainta, P.~Steinmetz, B.~Nestler, A.~Dennstedt, A.~Genau,
  M.~Bauer, H.~Köstler, U.~Rüde, Large scale phase-field simulations of
  directional ternary eutectic solidification, Acta Materialia 93 (2015) 194 --
  204.
\newblock \href {http://dx.doi.org/10.1016/j.actamat.2015.03.051}
  {\path{doi:10.1016/j.actamat.2015.03.051}}.

\bibitem{bauer15phasefield}
M.~Bauer, J.~H{\"o}tzer, M.~Jainta, P.~Steinmetz, M.~Berghoff, F.~Schornbaum,
  C.~Godenschwager, H.~K{\"o}stler, B.~Nestler, U.~R{\"u}de, Massively parallel
  phase-field simulations for ternary eutectic directional solidification, in:
  Proceedings of the International Conference for High Performance Computing,
  Networking, Storage and Analysis, ACM, 2015, p.~8.
\newblock \href {http://dx.doi.org/10.1145/2807591.2807662}
  {\path{doi:10.1145/2807591.2807662}}.

\end{thebibliography}

\end{document}